\documentclass{article}
\usepackage[utf8]{inputenc}
\usepackage[T1]{fontenc}
\usepackage{lmodern}
\usepackage[english]{babel}
\usepackage{amsmath, amsfonts, algorithm, algpseudocode, hyperref, geometry, amsthm, graphicx, booktabs, mathtools, setspace, diagbox,multirow, caption}
\usepackage{verbatim}
\usepackage[hang]{footmisc}
\newcommand{\indep}{\perp \!\!\! \perp}
\theoremstyle{plain}

\newtheorem{pro}{Proposition}

\title{Optimal execution in a fractional setting with reinforcement learning}
\author{Matthieu Garcin\textsuperscript{a,b} \and Jiang Pu\textsuperscript{a} \and Thomas Verrier\textsuperscript{a,c}}
\date{\today}

\begin{document}
\maketitle
{\renewcommand{\thefootnote}{\alph{footnote}}
\footnotetext[1]{De Vinci Higher Education, De Vinci Research Center, Paris, France.}
\footnotetext[2]{This research benefited from the support of the Chaire ``Deep Finance Statistics'' between QRT, Ecole Polytechnique and its foundation.}
\footnotetext[3]{CEREMADE, Université Paris Dauphine-PSL, Paris, France.}}
\begin{abstract}
    We consider the problem of optimal liquidation with linear market impact, as in the seminal work of Almgren and Chriss, but with an asset driven by a fractional Brownian motion. For $H \neq 1/2$, we observe that the deterministic solution given by Almgren and Chriss is no longer optimal. We develop a strategy to exploit the information induced by the correlation between price increments. The strategy is derived from a policy gradient algorithm, parameterized by a neural network, trained to maximise the CARA utility, for which our method allows for significant improvements.
\end{abstract}
\section{Introduction}

The optimal liquidation problem consists of selling a quantity $Q$ of shares from an asset of value $S$ over a time period $T$. A market impact phenomenon affects the price at which we sell as a linear function of our trading speed. In the original paper by Almgren and Chriss \cite{almgren2000}, the price follows a traditional Brownian motion, and one maximizes an objective function defined as a trade-off between the mean and the variance of $\Pi_T$, the final payoff: 

\begin{equation}
    \label{meanvar}
    \underset{v ,\,\, q_T = 0}{\operatorname{sup}}\operatorname{MV}_\lambda\{\Pi_T\} = \underset{v ,\,\, q_T = 0}{\operatorname{sup}}\left(\mathbb{E}\{\Pi_T\} - \lambda\operatorname{Var}\{\Pi_T\}\right),
\end{equation}

where $\Pi_T$ is a function of $v$, our liquidation speed process, $q_T = \int_0^T v_tdt$, the terminal inventory, and where $\lambda > 0$ is the risk-aversion parameter. It is worth noting that, since Almgren and Chriss only consider deterministic strategies, the variance can be computed in closed-form as $\Pi_T$ is a deterministic linear transformation of the Brownian motion. 

However if we consider dynamic strategies, even when restricting to the martingale case, the above mean-variance criterion can lead to time inconsistency, in the sense that after finding an optimal strategy $v^{t_0, x_0}$ for a time $t_0$ in the state $x_0$, an investor might want to diverge from this strategy at a future time $t$. This is caused by the second term in $\operatorname{Var}\{\Pi_T\}=\mathbb{E}\{\Pi_T^2\} - (\mathbb{E}\{\Pi_T\})^2$, which prevents us from applying Bellman's dynamic programming principle. This leads to formulating several definitions of optimality and the solution involves constrained optimization problems \cite{pedersen2017optimal, timeconsistency}.

The fractional Brownian motion (fBm) was introduced by Mandelbrot and van Ness \cite{mandelvanness} and proposed by Mandelbrot as a tool to model market dynamics \cite{mandelbrot}. The long- or short-range dependence of asset returns is a property originally motivating the use of fBm in finance. It remains empirically debated \cite{cont, famafrench, lo1991, willingeretal}. While the fBm can present arbitrage opportunities in continuous times \cite{rogers, guasoni2021high}, introducing a minimal time period between two trades prevents them \cite{cheridito2003arbitrage}. Arbitrages on a geometric fBm also vanish in continuous time once market frictions are introduced \cite{guasoni2006}.

Under this new fractional dynamic, the deterministic solution of Almgren and Chriss is no longer optimal. An explicit solution that maximizes only the gain expectation has been proposed \cite{dolinsky2024}. In detail, it consists in maximizing $\mathbb{E}\{\Pi_T\}$, for $S$ any right continuous price process with left limits such that $\mathbb{E}\left\{\int_0^T S_t^2 dt\right\} < \infty$. This result can be applied to the fBm case, which involves taking the conditional expectation of $\int_0^TS_tdt$, computed using the kernel representation  of the fBm \cite{dolinsky2024}. One can also have an explicit solution to the maximization of the gain expectation, with a square-integrable price process that does not need to be a semimartingale, and with a price impact formulated as the integral of a power function over the last trades \cite{Forde04032022}.

Alternatively, also for prices following an fBm, one can focus on the trade-off between the replication error of a perfect hedging strategy derived from a frictionless case and the trading cost \cite{bank2017hedging}. 

Another approach would be to use stochastic control techniques designed for systems driven by fBm, which lead to a maximum principle \cite{han2013maximum}. However the theory is restricted to the Hurst parameter $H \in (1/2, 1)$, our model lacks some of the required regularity conditions, and the framework does not allow for a terminal liquidation constraint. Most importantly, the maximum principle only yields a necessary condition of optimality, from which no explicit strategy can be easily extracted.

Our primary motivation is to bypass the absence of closed-form solutions for the optimal liquidation problem under fBm dynamics with a risk-sensitive objective such as formula \eqref{meanvar}. When $H \neq 1/2$, price increments exhibit serial correlation, so past prices carry information that a dynamic policy could, in principle, exploit. This raises two natural questions: can a learned policy conditioned on price history outperform the deterministic Almgren-Chriss schedule, and how does the potential gain depend on the Hurst exponent $H$?

In this article, we work on a probability space $(\Omega, \mathcal{G}, (\mathcal{G}_t)_{t \ge 0}, \mathbb{P})$, on which the fBm is defined, and we focus on another criterion instead of the mean-variance trade-off \eqref{meanvar}, namely the constant absolute risk aversion (CARA) utility, that is
\begin{equation}
    \label{CARA}
    \underset{v ,\,\, q_T = 0}{\operatorname{sup}}\mathbb{E}\{-\exp(-\lambda \Pi_T)\}\,,
\end{equation}
where the supremum is taken over the admissible set of strategies
\begin{eqnarray}
    \mathcal{A} = \left\{(v_t)_{t\in[0,T]}\in P(0,T),\int_0^T|v_t|dt\in L^\infty(\Omega)\right\},
\end{eqnarray}
where $P(0,T)$ denotes the set of progressively measurable processes defined on $[0,T]$.
To motivate this choice, we first mention that, for the traditional Brownian case, the CARA utility, unlike the mean-variance trade-off, allows one to apply Bellman's principle \cite{pham:hal-00401892} to derive the associated Hamilton-Jacobi-Bellman (HJB) equation. One can then solve the PDE and obtain the optimal control, which happens to be deterministic and a continuous version of the solution provided by Almgren and Chriss \cite{almgren2000}. We will use this policy as a benchmark in our experiments. Second, since for $H=1/2$ the optimal control is deterministic, optimizing over progressively measurable controls reduces to optimizing over deterministic ones. In that case, $\Pi_T$ is Gaussian and the exponential expectation in criterion \eqref{CARA} is evaluated through its moment-generating function, recovering the classical mean-variance trade-off, with $\tilde{\lambda} = \lambda/2$: 
\begin{equation*}
    \underset{v \, \text{deterministic},\, q_T = 0}{\operatorname{argmax}}\mathbb{E}\{-\exp(-\lambda \Pi_T)\}
    = \underset{v \, \text{deterministic},\, q_T = 0}{\operatorname{argmax}} \operatorname{MV}_{\tilde\lambda}\{\Pi_T\}\,.
\end{equation*}
Even if our strategies for $H \neq 1/2$ will not be deterministic, minimizing the exponential of $-\Pi_T$ means we have a great aversion to negative payoffs, and tends to compress the tails and reduce the variance of the payoff distribution, which we verify experimentally in our results.

We want to exploit the correlation between returns of the fBm, and thus use a strategy conditioned on past prices. However we do not have an explicit optimal solution so we aim for a numerical approximation. Our approach is to use the policy gradient algorithm \cite[Chap.~13]{suttonbarto}, parameterized with a neural network (NN). This algorithm is specifically designed to estimate optimal dynamic controls and is based on a stochastic gradient approximation.

We develop a policy gradient algorithm that learns an execution strategy conditioned on a forecast of future returns built from past price increments. The learned policy outperforms the Almgren-Chriss baseline for every $H \neq 1/2$ and recovers it in the martingale case. The improvement is larger when $H$ is far from $1/2$, with an asymmetry in favor of the anti-persistent regime. The policy also remains robust to estimation errors on $(H, \sigma)$.

The article is organized as follows. Section~2 introduces the fBm, some simulation and estimations techniques later used, and an example of the time inconsistency of the mean-variance criterion in presence of fBm trajectories. Section~3 presents the liquidation model, provides the explicit solution in the martingale case, and describes our numerical approximation method: the features fed to the network, the policy gradient theorem with the Almgren-Chriss baseline, and the resulting REINFORCE algorithm. Section~4 reports the numerical results, including the dependence of the performance on $H$ and our hyper-parameters, the link with the mean-variance criterion, and a robustness study to parameter misspecification. Section~5 concludes.

\section{Preliminaries on the fBm}
\subsection{The fBm and its main properties}

An fBm with Hurst index $H \in (0,1)$ is a continuous adapted centered Gaussian process $B^H = (B_t^H)_{t \ge 0}$, such that $B_0^H=0$ and, $\; \forall \; t,\,s \ge 0,$ 
$$
\mathbb{E}\big\{B_t^H B_s^H\big\} = \frac{1}{2}\left[t^{2H} + s^{2H} - |t-s|^{2H}\right]\,.
$$

For $H = 1/2$, we retrieve the classical Brownian motion. The fBm is self-similar, in the sense that $(B^H_{at})_{t \ge 0} \overset{\mathcal{L}}{=} |a|^H \left(B^H_t\right)_{t \ge 0}$ . It is a.s. $\alpha$-Hölder for all $0<\alpha<H$.

The fractional Gaussian noise (fGn) is defined as $R^{H,\Delta} = \left(B_{t+\Delta}^H - B_t^H\right)_{t\ge0}$. It is the increment process of the fBm, it is stationary and has an autocovariance function $\gamma^\Delta(t-s) = \mathbb{E}\left\{(B_{t+\Delta}^H - B_{t}^H)(B_{s+\Delta}^H - B_s^H)\right\}$ given by
\begin{equation}
    \label{eq:autocorel}
\gamma^\Delta(t-s)
= \frac{1}{2}\bigl(\,|t-s-\Delta|^{2H} - 2|t-s|^{2H} + |t-s+\Delta|^{2H}\,\bigr).
\end{equation}

The sign of $\gamma^\Delta$ is entirely determined by $H$: for $H < 1/2$, increments are negatively correlated (anti-persistent), while for $H > 1/2$, they are positively correlated (persistent).

\subsection{Simulation of the fBm}
\label{sub:simul}
We use the circulant embedding method introduced by Wood and Chan \cite{chanwood,coeurjollysimul}. To generate $n$ sample paths of length $N$, the standard Cholesky approach requires $O(N^3)$ for the factorization plus $O(n\,N^2)$ for the matrix product, which is prohibitive for the sample sizes we need ($n =10^6$, $N = 800$). The Wood-Chan method exploits the stationarity of fBm increments: their covariance matrix $\Sigma_R$ is Toeplitz, as seen in equation \eqref{eq:autocorel} and can be embedded in a circulant matrix $C$ of size $m = 2^{\lceil \log_2 2(N-1) \rceil}$. Since circulant matrices are diagonalised by the discrete Fourier transform (DFT) of their first row, which we denote by $\mathbf{c}$, the eigenvalues are computed once in $O(m \log m)$, and each trajectory is then obtained as
\[
  R^{H,\Delta} = \frac{1}{\sqrt{m}}\,
  \mathrm{DFT}\,\!\bigl(\sqrt{\mathbf{c}} \odot W\bigr),
\]
where $W$ is a complex normal random vector built with cost $O(m)$. The total cost is $O(n\, m \log m) = O(n\, N \log N)$, and the method is exact.

\subsection{Estimating the Hurst parameter and the volatility}
\label{sub:estim}
We follow the classical discrete variation estimator \cite{istas1997quadratic, coeurjollyestim}. Given an observed fBm path
$(\sigma B^H_{k/N})_{0 \le k \le N}$, we form dilated second-order
differences
\[
  V^{(m)}_i = \sigma\bigl(B^H_{i/N} - 2\,B^H_{(i-m)/N}
    + B^H_{(i-2m)/N}\bigr),
  \qquad
  S_N(m) = \frac{1}{N +1- 2m}\sum_{i=2m}^{N}\bigl|V^{(m)}_i\bigr|^2.
\]
By self-similarity, $\mathbb{E}\{S_N(m)\} = \sigma^2(4 - 2^{2H})\,
m^{2H} N^{-2H}$. An ordinary least squares regression of $\log S_N(m)$ on $\log m$ for
$m = 1, \ldots, M$ yields the estimator $\widetilde{H}, \widetilde{\sigma}$.

The key advantage of second-order differences over simple increments is that they destroy enough long-range dependence for $(\widetilde{H}, \widetilde{\sigma})$ to be consistent at the rate $\sqrt{N}$ and asymptotically normal for all $H \in (0,1)$, whereas first-order increments only achieve this for $H < 3/4$ \cite{coeurjollyestim}.

\subsection{Time inconsistency of the mean-variance criterion: a two-period example}
\label{sub:time_inconsistency}

We illustrate the time inconsistency mentioned in the introduction on the simplest possible case. Let $S = S_0 + \sigma B^H$,   and the two consecutive returns $R_1 = S_1 - S_0 $ and $R_2=S_{2}-S_1$. They are centered Gaussian variables with
$$
\operatorname{Var}\{R_1\} = \operatorname{Var}\{R_2\} = \sigma^2, \qquad
\operatorname{Corr}\{R_1, R_2\} = \rho = 2^{2H-1} - 1 \in \left(-\tfrac12, 1\right).
$$
An investor holds no position on $[0,1]$. This is optimal since $\mathbb{E}\{R_1\}=0$. He then holds a position $w_2$ on $[1,2]$, chosen after observing $R_1$. The payoff is $P = w_2 R_2$.

We first derive the optimal rule as seen from $t=1$. Given that conditionally on $\mathcal{G}_1$, $R_2 \sim \mathcal{N}\big(\rho R_1,\, \sigma^2(1-\rho^2)\big)$, maximizing $\mathbb{E}\{P \mid \mathcal{G}_1\} - \lambda \operatorname{Var}\{P \mid \mathcal{G}_1\}$ over $\mathcal{G}_1$-measurable $w_2$ yields the linear rule
\[w_2 = k_1 R_1, \quad k_1 = \frac{\rho}{2\lambda \sigma^2\,(1-\rho^2)}\;.\]
We then derive the optimal rule as seen from  $t = 0$. Among the linear controls $w_2 = k R_1$, for $k\in \mathbb{R}$, the mean-variance criterion evaluated at $t=0$ is, using Isserlis' theorem,
$$ J(k) := \mathbb{E}\{P\} - \lambda\operatorname{Var}\{P\} = k\,\rho \sigma^2 - \lambda k^2 \sigma^4 (1 + \rho^2),
\; \text{maximized at} \;
k_0 = \frac{\rho}{2\lambda \sigma^2\,(1+\rho^2)}.
$$
For $H \neq 1/2$, we have $k_1 \neq k_0$: the rule that is optimal at $t=0$ is no longer optimal at $t=1$, and the investor wants to take a larger position than planned. The gap comes from the decomposition
$$
\operatorname{Var}\{P\} = \mathbb{E}\big\{\operatorname{Var}\{P\mid\mathcal{G}_1\}\big\} + \operatorname{Var}\big\{\mathbb{E}\{P\mid\mathcal{G}_1\}\big\},
$$
whose second term is penalized at $t=0$ but not taken into account at $t=1$.

We evaluate the two rules at $t = 0$:
$$
J(k_0) = \frac{\rho^2}{4\lambda(1+\rho^2)} \,>\,
J(k_1) = \frac{\rho^2\,(1 - 3\rho^2)}{4\lambda\,(1-\rho^2)^2}\,,$$ as soon as $H\neq1/2$. An investor who does not commit to $k_0$ and re-optimizes at $t=1$ therefore does strictly worse than planned, which is what we call a time inconsistency. For $\rho > 1/\sqrt{3}$, that is
$$
H > H^* = \frac12\left(1 + \log_2\!\Big(1 + \frac{1}{\sqrt3}\Big)\right) \simeq 0.83\,,
$$
we even get $J(k_1) < 0$: the re-optimizing behavior is worse than not trading at all, and $J(k_1) \to -\infty$ as $H \to 1$. At $H = 1/2$, $\rho = 0$ and both rules vanish: the inconsistency is created by the memory of the fBm. By contrast, the CARA criterion \eqref{CARA} is an expected utility and therefore it satisfies the property $\mathbb E\big\{\mathbb{E}\{-\exp(-\lambda\Pi_T)\mid\mathcal{G}_t\}\big\}=\mathbb{E}\{-\exp(-\lambda\Pi_T)\big\}$, preventing time inconsistency.
\section{Method for optimal liquidation with fractional dynamic}

In this section we describe the model, the features, the policy gradient method, and the training algorithm used to learn a liquidation strategy under fBm dynamics. We stress that the entire pipeline is $H$-specific: the Hurst exponent must be known both to simulate the training trajectories and to compute forecasts, so a distinct model needs to be trained for each value of $H$.

\subsection{Model and objective}
\label{sub:perfACToH}
We use a setup very similar to the classical approach \cite{almgren2000}, with $q$ the number of shares held, such that $q_0 = Q$, $ q_T = 0$, $v = -\dot{q}$ the selling speed, so that $v > 0$ when we sell, $S = S_0+ \sigma  B^H$ the price process, and $\tilde S = S - \eta v$ the execution price. The parameters $\sigma $ and $\eta$ are respectively the daily arithmetic volatility and the temporary market impact parameter. We denote $X$ the wealth process, such that 

\begin{equation}
    \label{eq:richesse}
X_t = \int_0^t \tilde S_s v_s ds\ = \int_0^t \left(S_sv_s - \eta v_s^2\right)ds  , \quad t \in [0,T] \,.
\end{equation}

We use an implementation shortfall benchmark so our payoff writes $$\Pi_T = X_T - Q S_0\,,$$ and we recall that our objective is to solve the control problem defined by formula \eqref{CARA}.

For $H=1/2$, the classical theory of stochastic control allows us to formally derive an HJB equation, which in this case can be solved explicitly.
Defining $\kappa = (\lambda \sigma ^2/2\eta)^{1/2}$ and $\tau = 1/\kappa$ the characteristic time of our problem, the optimal control writes itself
\begin{equation}
    \label{eq:ControlAC}
    \begin{dcases}
    v_t^\mathrm{AC} = \kappa Q \,\frac{\cosh\big(\kappa(T-t)\big)}{\sinh(\kappa T)} \\
    q_t^\mathrm{AC} = Q \,\frac{\sinh\big(\kappa(T-t)\big)}{\sinh(\kappa T)}\,\;.
    \end{dcases}
\end{equation}

The first thing to notice is that the optimal control is deterministic, and has finite variation, with no term in $dB^\frac12_t$. It is more regular than could be expected from a typical stochastic control problem. We will refer to this strategy as the AC strategy in the rest of this article, as it is a continuous version of the Almgren and Chriss solution \cite{almgren2000}.

\begin{figure}[htbp]
    \centering
    \includegraphics[width=\textwidth]{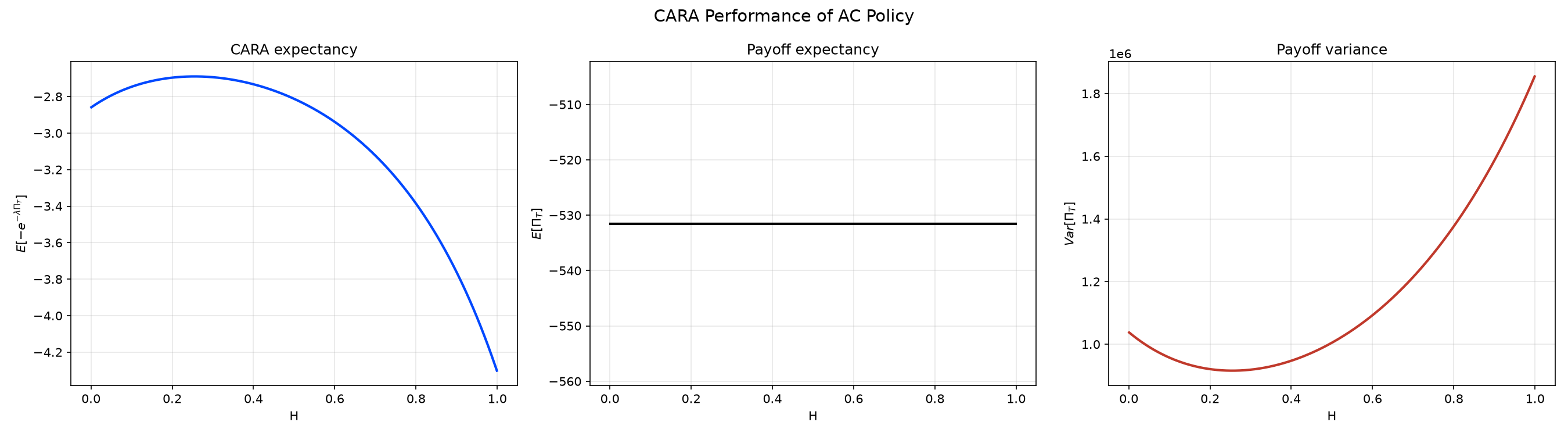}
    \captionsetup{width=0.7\textwidth}
    \caption{Analytical performance of the AC strategy under fBm as a function of the Hurst exponent $H$: expected CARA utility (left), expected payoff (center), and payoff variance (right).}
    \label{fig:perf_ac}
\end{figure}

Let us briefly see how the value of H impacts the performance of the deterministic strategy \eqref{eq:ControlAC}. In the case where $H\neq 1/2$, the payoff $\Pi_T$ of the AC strategy is (the a.s. - limit of) a linear transformation of the fBm and is therefore Gaussian, so the expected CARA utility follows in closed form from the Gaussian moment-generating function, $\mathbb{E}\{-\exp(-\lambda \Pi_T)\} = -\exp\!\big (-\lambda \operatorname{MV}_{\tilde\lambda}\{\Pi_T\}\big)$. Both the expectation and the variance are computed with the Fubini-Lebesgues theorem. It is apparent in Figure \ref{fig:perf_ac} that $\mathbb{E}\{\Pi_T\}$ is independent of the Hurst exponent, whereas $\mathrm{Var}\{\Pi_T\}$ increases with $H\to 0 \text{ or } 1$, with a minimum around $H\simeq0.25$. The asymmetry in favor of the anti-persistent regime can be linked to the fact that $\operatorname{Var}\left\{S_t-S_0\right\}=\sigma^2\,t^{2H}$ grows with $H$ for $t\ge1$. Thus the entire dependence of the utility on $H$ for the AC strategy is driven by the execution risk. We will see that our algorithm, which maximizes the CARA expectation, tends to reduce $\Pi_T$'s variance, even if it sometimes lowers its expectation. In that sense, it provides us with an indirect way of improving the mean-variance criterion \eqref{meanvar}.

\subsection{Features and information}
\label{sub:features}
To exploit the correlation between increments of the fBm, we want to condition our action at time $t$ on a vector of features named $F_t \in \mathbb{R}^p$. We want $F_t$ to summarize the current and past state of the system or, in other words, the information used to predict the best inventory at the next state. We choose to take $F_t=(t,\,q_t,\,S_t,\, \hat{R}_{t,t+\tau})^\top$. The time $t$ and inventory $q_t$ are the only variables used in the AC control, which makes them the natural minimal state. In addition to that, we consider as features $S_t$, the current price, and $\hat{R}_{t,t+\tau}$, which is the conditional expectation of the future price return over a horizon $\tau$, the characteristic time of our problem, which is present in the AC control \eqref{eq:ControlAC}. We write $\hat{R}_{t,t+\tau}=\mathbb{E}\{R_{t,t+\tau}\mid Z_t\}$, where $R_{u,v} = S_v - S_u$ and $Z_t=(R_{t-\delta_1,t-\delta_0},\dots,R_{t-\delta_p,t-\delta_{p-1}})^\top$. Here, $Z_t$ gathers $p$ past returns between lags $0 = \delta_0 < \delta_1<\dots<\delta_p$ . The forecast $\hat{R}_{t,t+\tau}$ summarizes the non-Markovian information carried by the fBm path into a single scalar, and vanishes to $0$ at $H=1/2$. Although the price level does not appear in the AC control, it carries information complementary to the predicted return. The first thing to note is that $F_t$ is $\mathcal{G}_t$-measurable, since we only look at the present state and at some past returns.  

We note that the conditional expectations of future returns of the fBm have already been used as trading signals, to maximize a time-additive mean-variance criterion, that is $\mathbb{E}\left\{\sum_{k=0}^{n-1} \operatorname{MV}_\lambda\{X_{t_{k+1}} - X_{t_{k}} \mid \mathcal{G}_{t_k}\}\right\}$ \cite{guasoni2021high}. We also confirm experimentally that the learning is faster and more stable compared to only giving $(t,q_t,S_t,Z_t)$ as an input to the NN. 

We want to draw attention to the fact that one needs to know the value of $H$ to compute $\hat{R}_{t,t+\tau}$. Therefore, the accuracy of the estimation of the Hurst parameter is of major importance.

Given the Gaussian nature of the fBm, the forecast $\hat{R}_{t,t+\tau}=\beta^\top Z_t$ is the standard covariance-based predictor: $\beta=\Sigma_Z^{-1}\Sigma_{RZ}^\top$ where $\Sigma_Z=\operatorname{Var}\{Z_0\}$, $\Sigma_{RZ} = \operatorname{Cov}\{R_{0,\tau}\,,Z_0\}$. Note that the covariance matrixes are identical for all values of $t$ since the fGn is stationary. The lags in $Z_t$ are chosen to maximize the mean directional accuracy, that is the probability that the predictor recovers the correct sign of the future return. Its explicit value is \cite[Thm.1]{garcinforecast}: 
\begin{equation}
\label{hit_ratio}
    r=\mathbb{P}[\,\hat{R}_{t,t+\tau}\,R_{t,t+\tau}\ge 0\,] = 1 - \frac1\pi \arctan \left( \sqrt{\frac{ \sigma ^2\tau^{2H}}{\Sigma_{RZ}\Sigma_Z^{-1}\Sigma_{RZ}^\top}-1}\right)\,.
\end{equation}

It exceeds $1/2$ as soon as $H\neq1/2$ and equals $1/2$ in the martingale case. Maximizing $r$ is natural for our liquidation problem: our policy needs the direction of the forthcoming price move to decide whether to trade ahead of or behind the AC schedule. The optimal lags are computed via numerical optimization given equation  \eqref{hit_ratio}, and depend only on $H$. With an odd number of lags, we get one lag equal to $\tau$, $(p-1)/2$ lags smaller than $\tau$ and $(p-1)/2$ lags larger than $\tau$ \cite{garcinforecast}. We also note that with $H$ fixed, if $(\delta_0,\dots,\delta_p)$ is optimal for a horizon of $1$, by self-similarity $(\tau\,\delta_0,\dots,\tau\,\delta_p)$ is optimal for a horizon of $\tau$. The number of lags $p$ is a hyperparameter controlling the depth of memory fed to the forecast.

\begin{figure}[htbp]
    \centering
    \includegraphics[width=0.8\linewidth]{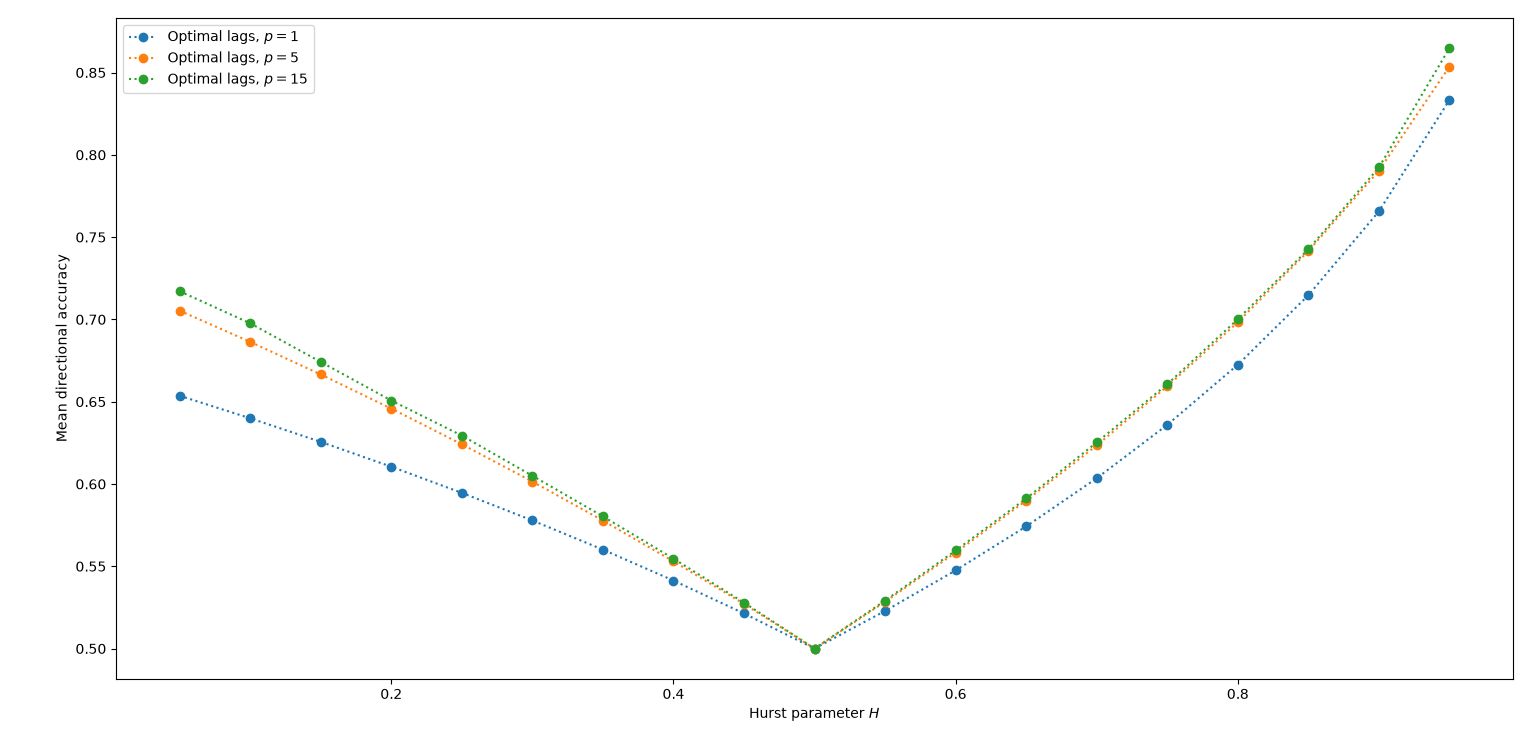}
    \captionsetup{width=0.7\textwidth}

    \caption{Mean directional accuracy $r$ for $p\in\{1,5,15\}$ optimal lags.}
    \label{fig:hit_ratio_values}
\end{figure}

Figure \ref{fig:hit_ratio_values} illustrates that $p=5$ is a good balance between computational efficiency and performance, as increasing $p$ beyond 5 only allows for relatively small improvements, both numerically and theoretically \cite{BROUTY2024114543}. We also note that, as usual, the more $H$ diverges from $1/2$, the more information we can extract from the returns of the fBm.

We also verify that the mean directional accuracy with $p=5$ optimal lags is higher than the one with $p=15$ evenly spread lags: $\delta_0=0,\, \delta_{k}=\tau (2k-1)/{p} \text{ for } k=1,\dots,p$, ensuring that our method allows for good performance while reducing computational cost. Indeed, to compute $\hat{R}_{t,t+\tau}$, one needs to invert a $p\times p$ matrix, and the size of the deepest lag increases with $p$, meaning that we have to simulate the fBm on a larger interval.

\subsection{The algorithm and the theory behind}
Our CARA criterion \eqref{CARA} is a stochastic control problem and we can get an estimation of its optimal value with a policy gradient method \cite[Chap.~13]{suttonbarto}. The policy gradient algorithm is designed for a discrete-time problem so we divide $[0,T]$ into $N$ equal intervals delimited by times $t_k = kT/N\, ,\; k \in \left\{0, \dots, N\right\}$.

Let \begin{equation*}
\pi_\theta(q \mid F) = \mathbb{P}^\theta\left(q_{t+1} = q \mid F_t = F \right)
\end{equation*}
be a probability distribution on the space of control $q \in \mathbb{R}$ conditioned on the features vector $F \in \mathbb{R}^p$. We call $\pi_\theta$ our policy and it is parameterized by a vector $\theta \in \mathbb{R}^d$. In our case, \begin{equation*}\pi_\theta(.\,|\,F) = \mathcal{N}\bigl(\mu_\theta(F), \,\sigma_{\mathrm{explo}}^2\bigr)\,,\end{equation*} with $\sigma_{\mathrm{explo}}$ being a fixed exploration parameter, and $\mu_\theta(F) \in \mathbb{R}$ being a multilayer perceptron with two hidden layers of 20 ReLU units, with weight $\theta$, taking the features $F$ as an input. 

The base of our policy, as provided in Algorithm \ref{alg:rollout} is the following: given an fBm path simulated with the Wood-Chan method described in Section \eqref{sub:simul}, we iterate over the time grid to produce an execution trajectory. We denote ``$X$ is $\mathcal{T}$-measurable '' by $X\,\underline{\in}\, \mathcal{T}$.

\begin{algorithm}[H]
\caption{Policy roll-out (exploration phase)}
\label{alg:rollout}
\begin{spacing}{1.3}
\begin{algorithmic}[1]
\Require fBm path $(B^H_{k\Delta})_{k}$, parameters $\theta$,
         exploration $\sigma_{\mathrm{explo}}$
\State $q_0 \gets Q$
\For{$k = 0, \ldots, N-2$}
    \State Observe and compute features $F_{t_k} = (t_k,\, q_{t_k},\, S_{t_k},\,
      \hat{R}_{t_k, t_k+\tau})$ \Comment{$F_{t_k}\, \underline{\in} \,\mathcal{G}_{t_k}$}
    \State $\mu_k \gets \mu_\theta(F_{t_k})$
    \State Draw $\varepsilon_{k+1} \sim \mathcal{N}(0,1)$, $\varepsilon_{k+1} \,\underline{\in}\, \mathcal{G}_{t_{k+1}}$, 
     $\varepsilon_{k+1} \indep \mathcal{G}_{t_{k}}$
    \State $q_{t_{k+1}} \gets \mu_k
      + \sigma_{\mathrm{explo}}\,\varepsilon_{k+1}$ \Comment{$q_{t_{k+1}} \, \underline{\in}\,\mathcal{G}_{t_{k+1}}$}
\EndFor
\State $q_{t_N} \gets 0$ \Comment{forced liquidation}
\State Compute $\Pi_T$ as a discretization of equation \eqref{eq:richesse}
\end{algorithmic}
\end{spacing}
\end{algorithm}

Each action $q_{t_{k+1}}$ is drawn from $\mathcal{N}(\mu_\theta(F_{t_k}),\, \sigma_{\mathrm{explo}}^2)$, where the noise $\varepsilon_{k+1}$ is independent from $\mathcal{G}_{t_k}$. This independence is essential for the policy gradient theorem to hold. At the last time step we force $q_{t_N} = 0$, ensuring that only fully liquidated strategies are evaluated.

Define $$
\begin{cases}
G^\mathrm{ac} = - \exp\left(-\lambda\Pi_T^\mathrm{ac}\right)\,\\
G^\mathrm{nn} = - \exp\left(-\lambda\Pi_T^\mathrm{nn}\right)\,\\
 J(\theta)=\mathbb{E}_{\pi_\theta}\left\{G^\mathrm{nn}\right\}\,,
\end{cases}
$$
so that $G^\mathrm{ac}$ is the CARA utility of the payoff of the AC policy given by equation \eqref{eq:ControlAC} and $J(\theta)$ is the expectation of the CARA utility given we followed the NN policy described in Algorithm \ref{alg:rollout} parameterized by $\theta$. As an important result of reinforcement learning, we have \cite[Chap.~13]{suttonbarto}  
\begin{equation}  
\label{policyGrad}
\begin{split}
\nabla_\theta J(\theta) 
&\propto \sum_{k=0}^{N-2} \mathbb{E}_{\pi_\theta}\!\bigl\{\,G^\mathrm{nn}\, \,\nabla_\theta\log \pi_\theta(q_{t_{k+1}}|F_{t_k})\,\bigr\} \\
&= \sum_{k=0}^{N-2}  \,\mathbb{E}_{\pi_\theta}\left\{G^\mathrm{nn}\, \,\frac{q_{t_{k+1}} - \mu_\theta(F_{t_k})}{\sigma_{\mathrm{explo}}^2} \nabla_\theta \mu_\theta(F_{t_k})\right\} \\
&= \sum_{k=0}^{N-2}  \,\mathbb{E}_{\pi_\theta}\bigl\{G^\mathrm{nn}\, \,\varepsilon_{k+1}\, \nabla_\theta \mu_\theta(F_{t_k})\bigr\}\;.
\end{split}
\end{equation}

The second line follows from the log-likelihood of the Gaussian exploration policy $\pi_\theta(\cdot|F_{t_k}) = \mathcal{N}(\mu_\theta(F_{t_k}),\,\sigma_{\mathrm{explo}}^2)$. Each term in the sum has a clear role. First, the reward $G^\mathrm{nn}$ weights the update: actions that led to higher reward are reinforced more strongly. Second, the score of the exploration policy $\nabla_\theta \log \pi_\theta(q_{t_{k+1}}|F_{t_k})$ points in the parameter direction that makes the sampled action $q_{t_{k+1}}$ more likely under $\pi_\theta$.

Since we enforce $q_{t_N}=0$, the sum in equation \eqref{policyGrad} stops at $G^\mathrm{nn} \;\nabla_\theta\log\pi_\theta(q_{t_{N-1}} \mid F_{t_{N-2}})$.

The REINFORCE algorithm \cite[Chap.~13]{suttonbarto} is a stochastic gradient ascent method. Given equation \eqref{policyGrad}, we know that our gradient writes itself as the expectation of a random variable depending on the policy that we learned. We will then approximate this gradient by a batch of $m$ i.i.d. realizations of this random variable, that is

\begin{equation*}
    \widehat{\nabla_\theta J}(\theta) = \frac1m\sum_{i=1}^m \, \sum_{k=0}^{N-2} G^{\mathrm{nn},i}\,\, \frac{q_{t_{k+1}}^i - \mu_\theta(F_{t_k}^i)}{\sigma_{\mathrm{explo}}^2} \nabla_\theta \mu_\theta(F_{t_k}^i)\,.
\end{equation*}

To reduce the variance of the algorithm, we compare the learned policy to a baseline one \cite[Chap.~13]{suttonbarto}. For this, we need an approximation of the value awarded by our policy. We take our baseline as the AC strategy given by equation \eqref{eq:ControlAC}. Both policies are evaluated on the same fBm paths. The temporal advantage is then defined as
$$
  A_{t_k}^i \;=\; G^{\mathrm{nn},i}
        \;-\; \mathbb{E}\!\left\{G^{\mathrm{ac}, i}
        \;\middle | \;\mathcal{G}_{t_k}\right\}, \quad i \in \left\{1, \dots, m\right\},\, k\in \left\{1, \dots, N\right\}\,.
$$

The final payoff $\Pi_T^\mathrm{ac}$ is Gaussian conditionally on $\mathcal{G}_{t_k}$, so $\mathbb{E}\!\left\{G^{\mathrm{ac}, i}
        \;\middle | \;\mathcal{G}_{t_k}\right\}$ is computed as the moment-generating function of a Gaussian law. Both rewards must be computed on the same Brownian trajectory to reduce the variance. The modified gradient approximation is then
\begin{equation}
    \label{gradApproxBaseline}
    \widetilde{\nabla_\theta J}(\theta) = \frac1m\sum_{i=1}^m \sum_{k=0}^{N-2} \, A_{t_k}^i\,\frac{q_{t_{k+1}}^i - \mu_\theta(F_{t_k}^i)}{\sigma_{\mathrm{explo}}^2} \, \nabla_\theta \mu_\theta(F_{t_k}^i)\,.
\end{equation}

From this expression we understand that the greater $A_{t_k}^i$, that is the better our policy performed against the AC baseline, the more $\mu_\theta$ is pushed toward the sampled action. Averaging over a batch is essential: a single trajectory may favor the learned policy by chance (e.g. a sustained price trend), and only the mean advantage across diverse paths reveals whether the deviation from AC is genuinely beneficial.

Using basic properties of the conditional expectation and the independence between $\left(q_{t_{k+1}} - \mu_\theta(F_{t_k})\right)/\sigma_\mathrm{explo} = \varepsilon_{k+1} $, the exploration noise, and $\mathcal{G}_{t_k}$, we get

\begin{equation}
\label{nullBias}
\begin{aligned}
\mathbb{E}&\!\left\{
    \mathbb{E}\!\left\{G^{\mathrm{ac}}
      \,\middle|\, \mathcal{G}_{t_k}\right\}
    \frac{q_{t_{k+1}} - \mu_\theta(F_{t_k})}{\sigma_{\mathrm{explo}}^2}
    \,\nabla_\theta \mu_\theta(F_{t_k})
  \;\middle|\; \mathcal{G}_{t_k}\right\} \\
\;
  &= \mathbb{E}\!\left\{G^{\mathrm{ac}}
    \,\middle|\, \mathcal{G}_{t_k}\right\}
  \nabla_\theta \mu_\theta(F_{t_k})
  \;\underbrace{\mathbb{E}\!\left\{
    \frac{q_{t_{k+1}} - \mu_\theta(F_{t_k})}{\sigma_{\mathrm{explo}}^2}
  \;\middle|\; \mathcal{G}_{t_k}\right\}}_{= \,0} \\
  &=0\,.
\end{aligned}
\end{equation}

From equations \eqref{policyGrad} and \eqref{nullBias}, we conclude that 

\begin{equation*}
    \nabla_\theta J(\theta) \propto \mathbb{E}\left\{\widehat{\nabla_\theta J}(\theta)\right\} = \mathbb{E}\left\{\widetilde{\nabla_\theta J}(\theta)\right\}\,,
\end{equation*} 
proving that the modified gradient approximation is non biased.

In the gradient approximation \eqref{gradApproxBaseline}, all terms except $\nabla_\theta \mu_\theta(F_{t_k}^i)$ are obtained by simulation and simple computation. The gradient $\nabla_\theta \mu_\theta$ is computed by automatic differentiation using PyTorch's \texttt{backward} and \texttt{grad} command. Therefore, during the learning phase, after rolling our policy on a batch of $m$ fBm trajectories, we update our NN parameters with $\theta \gets \theta + \alpha\widetilde{\nabla_\theta J}(\theta)$, with $\alpha$ the learning rate parameter and keep iterating until the performance stops improving significantly.

\section{Numerical results}

We now study the performance of our strategy on simulations. We first detail the parameters used. Then, we display the results and compare them to the AC case, both in terms of CARA utility and mean-variance criterion, trying to explain the rationale of the policy, depending on the value of $H$. We also discuss the robustness of our study by providing confidence bounds to our estimators of the performance and by assessing the impact of estimating the parameters on a limited dataset.

\subsection{Choice of parameters} 
We follow the calibration reasoning of Almgren and Chriss, who derive orders of magnitude for the impact and risk-aversion parameters from realistic market quantities, namely price, daily volume, and volatility \cite{almgren2000}. We adopt the following values:

\begin{center}
\begin{tabular}{r@{:\;}l@{\;=\;}l}
Initial price of the asset & $S_0$ & $100 \;\$/\text{share}$ \\
Quantity to be liquidated & $Q$ & $1000$ shares \\
Liquidation horizon & $T$ & $5$ days \\
Arithmetic daily volatility & $\sigma$ & $0.015\,S_0 \;\;(\$/\text{share})/\text{day}^{1/2}$ \\
Temporary market impact coefficient & $\eta$ & $10^{-3} \;\; \$\times\text{day}/\text{share}^2$ \\
Risk-aversion parameter & $\lambda$ & $10^{-3} \;\; \$^{-1}$ \\
\end{tabular}
\end{center}

These values lead to our characteristic time $\tau = (2\eta/\lambda \sigma ^2)^{1/2} \simeq 0.94$ days, roughly $20\%$ of our liquidation period. The choice of the discretization step $\Delta$ impacts the computing cost and the precision of the algorithm. We choose to take $\Delta= 0.05$ days, that is $1\%$ of T to correctly approximate the integral from $0$ to $T$ in \eqref{eq:richesse}. However, the forecast $\hat{R}_{t,t+\tau}$ requires price increments reaching back to $\delta_p$ days before the start of liquidation: for $p = 5$ the deepest lag across the values of $H$ is approximately $35$ days. We therefore simulate each fBm trajectory on the extended window $[-L,\, T]$ with $L = 35$, bringing the total number of steps to $N=(T + L)/\Delta = 800$.

\subsection{Statistical analysis of the Monte Carlo estimator}

To give a monetary meaning to the CARA utility, we will often use the certainty equivalent:
\begin{equation*}
    \operatorname{CE} = g\bigl(\mathbb{E}\!\left\{-e^{-\lambda\Pi_T}\right\}\bigr) \text{ with } g(x)= -\frac{1}{\lambda}\log(-x)\,.
\end{equation*}
It is the deterministic payoff that yields the same expected utility as the random outcome $\Pi_T$. The $\operatorname{CE}$ metric helps to assign a monetary meaning to the CARA utility, and allows for a clearer understanding of the values.

During the test phase, we use the same policy rollout as in Algorithm \ref{alg:rollout}, except that we do not add an exploration noise to the output of the NN, and simply take $q_{t_{k+1}}=\mu_\theta\left(F_{t_k}\right)$. Each simulated trajectory $i$ produces a payoff $\Pi_T^i$ that is a deterministic measurable functional of the fBm path $B^{H,i}$. Since the paths are drawn independently across trajectories, the CARA utilities $G^i = -\exp(-\lambda\Pi_T^i)$ are i.i.d.

We want to build confidence intervals for the CE metrics obtained by simulations. In order to use the central limit theorem, we impose a slightly stronger admissibility condition in this paragraph, which restricts the set of admissible strategies as detailed in Proposition~\ref{pro:sq_int}. In practice, this constraint can be enforced by truncating the output of the ANN. In all our numerical experiments, however, such truncation is never triggered.

\begin{pro}\label{pro:sq_int}
Restricting the set of admissible strategies to those satisfying $\forall t\in[0,T], |v_t|\le C$ a.s. for some fixed constant $C>0$, the $G^i$s are square integrable.
\end{pro}

\begin{proof}
For notational simplicity, we drop the superscript $i$. We have
\begin{eqnarray*}
\mathbb{E}[G^{2}] & = & \mathbb{E}[\exp(-2\lambda\left(X_{T}-QS_{0}\right))].
\end{eqnarray*}

Using the boundedness of $v$, we obtain
\begin{eqnarray*}
\left|X_{T}\right| & \le & \int_{0}^{T}\left|S_{t}v_{t}\right|dt+\eta\int_{0}^{T}v_{t}^{2}dt\\
 & \le & CT\sup_{t\in[0,T]}\left|S_{t}\right|+\eta C^{2}T.
\end{eqnarray*}
Therefore, it suffices to show that for every $z\ge0,$ the Laplace transform $\mathbb{E}\left[\exp\left(z\cdot\sup_{t\in[0,T]}\left|S_{t}\right|\right)\right]$
is finite.
By Borell-TIS inequality, there exist constants \(C_1,C_2>0\) such that \[ \mathbb{P}\!\left( \sup_{t\in[0,T]} |S_t| \ge x \right) \le C_1 e^{-C_2 x^2}, \qquad x\ge 0. \] Hence \(\sup_{t\in[0,T]} |S_t|\) has Gaussian tails and therefore admits exponential moments of all orders.
\end{proof}

We denote $\mu = \mathbb{E}\{G^i\}$ and $V = \operatorname{Var}\{G^i\}$. By the central limit theorem,
$$
  \sqrt{n}\,(\bar{G}_n - \mu)
  \;\xrightarrow{\mathcal{L}}\;
  \mathcal{N}(0,\,V)\,, \text{ where } \bar G_n = \frac1n\sum_{i=1}^nG^i\,.
$$
The certainty equivalent is $\operatorname{CE} = g(\mu)$. Therefore, by the delta method and $g'(x) = -{1}/\lambda\, x$, we have
$$
  \sqrt{n}\,\bigl(g(\bar{G}_n) - \operatorname{CE}\bigr)
  \;\xrightarrow{\mathcal{L}}\;
  \mathcal{N}\!\left(0,\;\frac{V}{\lambda^2\mu^2}\right)\,.
$$
By Slutsky's lemma and the continuous mapping theorem ($x \mapsto \, | x |$ is continuous), replacing $V$ and $\mu$ by their empirical counterparts $\hat V_n$ and $\bar{G}_n$ preserves the convergence, yielding the asymptotic $95\%$ confidence interval 
\begin{equation*}
  \widehat{\operatorname{CE}}_n
  \;\pm\;
  1.96\;\frac{\sqrt{\hat{V}_n}}{\lambda\,|\bar{G}_n|\,\sqrt{n}}.
\end{equation*}

\subsection{First results}

As mentioned in Section~\ref{sub:features}, the choice of $p$, the number of lagged returns, is a balance between the amount of information extracted from past increments and the computational cost: each additional lag requires inverting a larger $p \times p$ covariance matrix and extends the warm-up window $L$, increasing the length of the simulated trajectories. Table~\ref{tab:Pchoice} shows that, for $H = 0.2$, most of the gain over AC is already captured at $p = 5$, with diminishing returns beyond. For other values of $H$, the deepest lag for $p=7$ is larger than $100$ days, making the simulation significantly more expensive for a marginal improvement.

\begin{table}[H]
    \centering
    \begin{tabular}{c|ccccc}
    \toprule
    Strategy  & AC  &  NN $(p=1)$   & NN $(p=3)$  &NN $(p=5)$&NN $(p=7)$ \\
    \midrule
    $\operatorname{CE}$   & $-991.8$  & $-699.6 \pm 3.4$ & $-662.3 \pm 3.7 $&$ -644.7 \pm 3.0$&$ -638.4 \pm 3.8$ \\
    \bottomrule
    \end{tabular}
    \captionsetup{width=0.7\textwidth}
    \caption{Certainty equivalent of the learned policy for increasing
    numbers of lagged returns $p$, evaluated at $H = 0.2$ ($n = 10^6$
    trajectories, $95\%$ confidence intervals). The AC baseline uses no
    price history.}
    \label{tab:Pchoice}
\end{table}

In order to check the relevance of the learned strategy with respect to $H$, we report in Table~\ref{tab:cross_eval} the CE performance of three policies, each one trained on a given $H$ value and evaluated on test trajectories of various Hurst exponents. The expected pattern holds: for each value of $H\neq1/2$, the network trained on the real value of the Hurst exponent performs best (the diagonal dominates each column). In particular, the strategy that exploits serial dependence outperforms AC as soon as $H\neq1/2$. For $H=1/2$, where the price is a martingale and no forecast is exploitable, we recover the AC value.

\begin{table}[H]
    \centering
    \begin{tabular}{c|ccc}
        \toprule
        \diagbox{Strategy}{Test} &  $H=0.35$ &  $H=0.5$ &  $H=0.65$ \\
        \midrule
        NN ($H=0.35$) & $\mathbf{-900.4 \pm 3.0}$ & $-1062.0 \pm 3.4$ & $-1115.9 \pm 3.2$ \\
        NN ($H=0.5$)  & $-1014.3 \pm 2.4$ & $\mathbf{-1035.1 \pm 2.5}$ & $-1089.6 \pm 2.7$ \\
        AC            & $-996.3$ & $\mathbf{-1034.0}$ & $-1106.1$ \\

        NN ($H=0.65$) & $-1037.8 \pm 2.4$ & $-1072.6 \pm 2.4$ & $\mathbf{-1035.0 \pm 2.3}$ \\
        \bottomrule
    \end{tabular}
    \captionsetup{width=0.7\textwidth}
    \caption{Cross-evaluation of the certainty equivalent (higher is better), with $95\%$ confidence intervals ($n = 10^6$ trajectories, $p=5$). Bold entries mark the best strategy per column.}
    \label{tab:cross_eval}
\end{table}
\subsection{Visualization of the policies}
\label{sub:vizu}
To understand in detail how our strategy outperforms the AC baseline beyond the aggregated results of Table \ref{tab:cross_eval}, we fix $H = 0.7$ and $0.3$ and display three test trajectories. We focus on the trajectories corresponding to the median, the first percentile, and the $99$th percentile of the criterion $\Pi_T^\mathrm{nn} - \Pi_T^\mathrm{ac}$. For each of them we illustrate the price path $S$ and the inventory trajectories of both strategies.

\subsubsection{Persistent regime (\texorpdfstring{$H = 0.7$}{H = 0.7})}

\begin{figure}[htbp]
    \centering
    \includegraphics[width=0.85\linewidth]{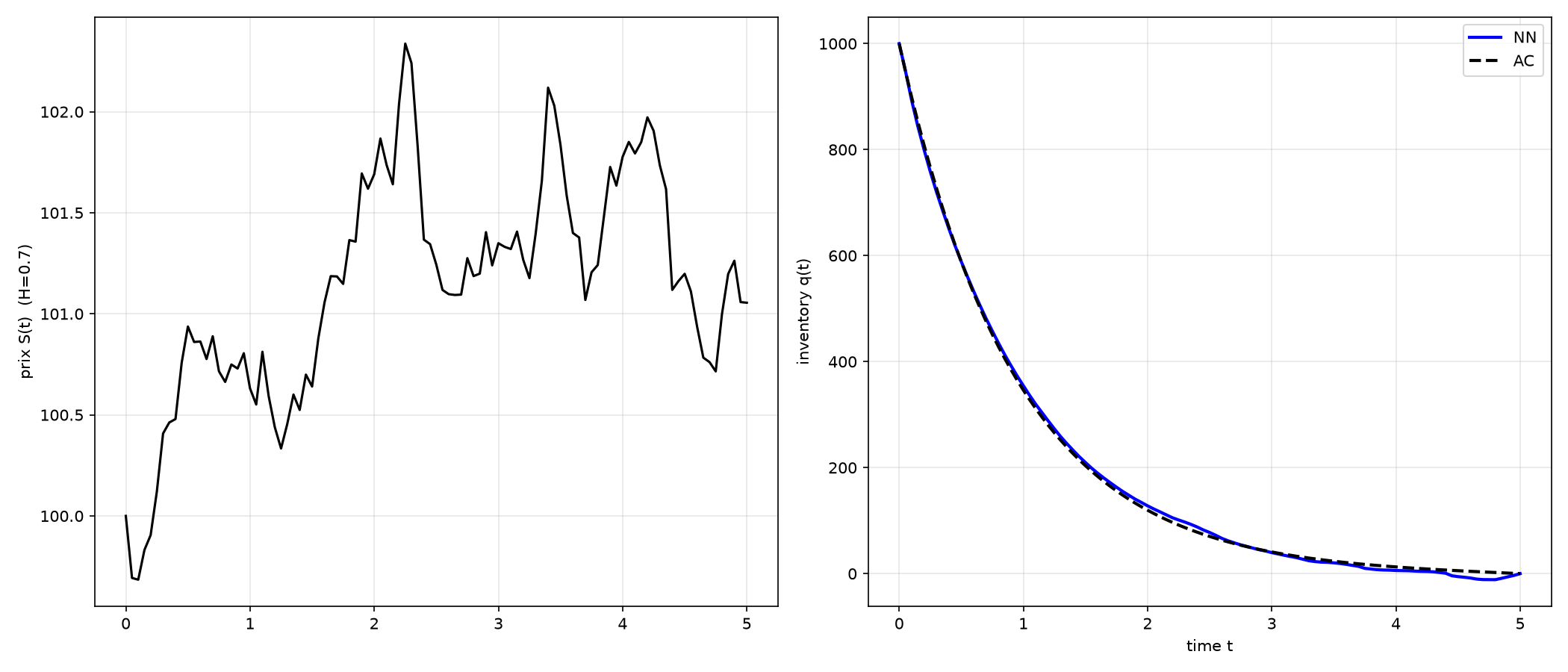}
    \captionsetup{width=0.7\textwidth}
    \caption{Median trajectory at $H=0.7$ ($\Pi_T^\mathrm{nn} - \Pi_T^\mathrm{ac}=-10$).}
    \label{fig:med07}
\end{figure}

One of the main feature of an fBm in the persistent regime is the possible presence of a price trend caused by the positive correlation between increments. For the median trajectory, we see a Brownian path without a clear trend (Figure~\ref{fig:med07}) and the two inventory curves are almost indistinguishable: the network treats the AC schedule as its default and departs from it only marginally; we observe a slight loss in performance.

\begin{figure}[htbp]
    \centering
    \includegraphics[width=0.85\linewidth]{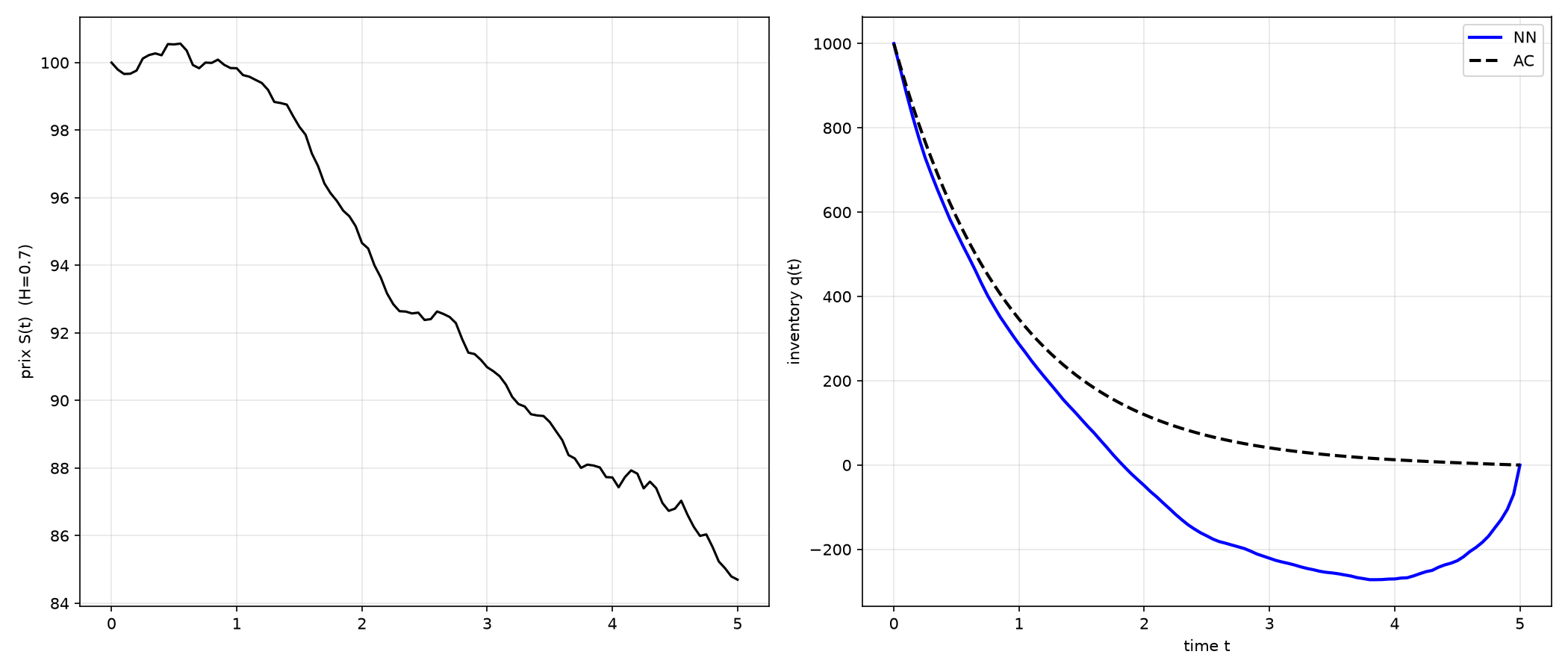}
    \captionsetup{width=0.8\textwidth}
    \caption{$99$th percentile trajectory at $H=0.7$ ($\Pi_T^\mathrm{nn} - \Pi_T^\mathrm{ac} = +2483$).}
    \label{fig:perc99_07}
\end{figure}

\begin{figure}[htbp]
    \centering
    \includegraphics[width=0.85\linewidth]{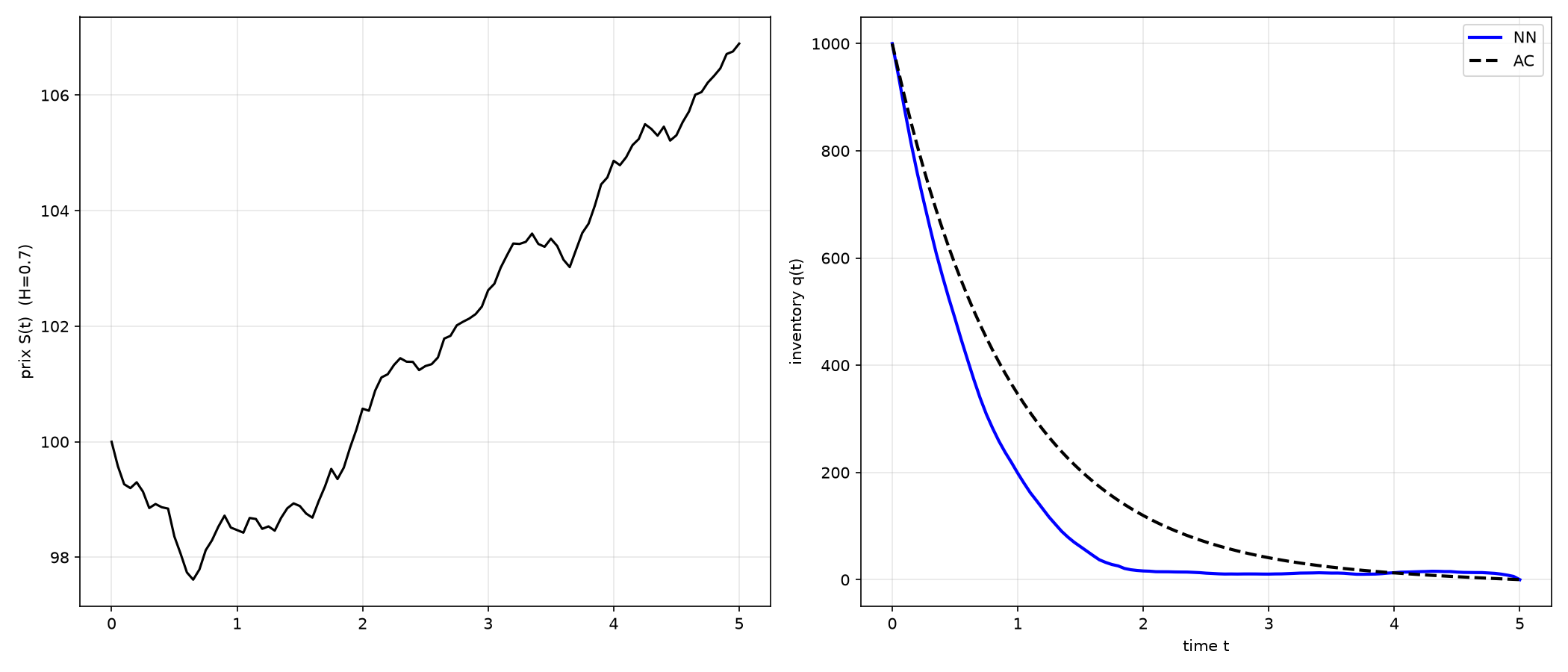}
    \captionsetup{width=0.8\textwidth}
    \caption{$1$st percentile trajectory at $H=0.7$ ($\Pi_T^\mathrm{nn} - \Pi_T^\mathrm{ac} = -569$).}
    \label{fig:perc1_07}
\end{figure}

Figure~\ref{fig:perc99_07} shows where the CARA expected performance is earned. The price falls steadily from $100$ to $85$, and persistence means that, once the move starts, the forecast keeps predicting further decreases. The network liquidates well ahead of schedule, drives its inventory to $-270$ shares, and buys them back at the very end at a much lower price, outperforming AC. Figure~\ref{fig:perc1_07} is the symmetric failure: the price drifts down over the first day, the forecast extrapolates, the network is essentially fully liquidated by $t=2$, but the trend reverses and the price climbs back. The shares were sold too soon, and AC outperforms the NN strategy. In both cases the mechanism is the same: the network bets on the continuation of an observed trend, but the magnitudes are asymmetric: the $99$th percentile gains more than four times what the first percentile loses. 

By going short, the network reverses its exposure to price movements: a further price decrease now becomes a gain, since the shares are repurchased more cheaply. This admits a natural interpretation in terms of hedging. In the multi-asset problem, a trader liquidating one asset may find it optimal to take a temporary position in a second, correlated one: this hedging position offsets part of the price risk carried by the inventory, and is unwound by $T$ \cite{gueant2015convex}. Our setting contains no second asset, but the serial correlation of the fBm increments plays the same role across time rather than across assets. The asset partially hedges itself, and the short position is a round trip that is opened once the inventory is exhausted and closed before the terminal date.

For {persistent regimes} ($H > 1/2$), the average out-performance on the CARA utility {is not obtained by doing better on most paths}, but {rather on the extreme cases}.

\subsubsection{Anti-persistent regime (\texorpdfstring{$H = 0.3$}{H = 0.3})}

\begin{figure}[htbp]
    \centering
    \includegraphics[width=0.85\linewidth]{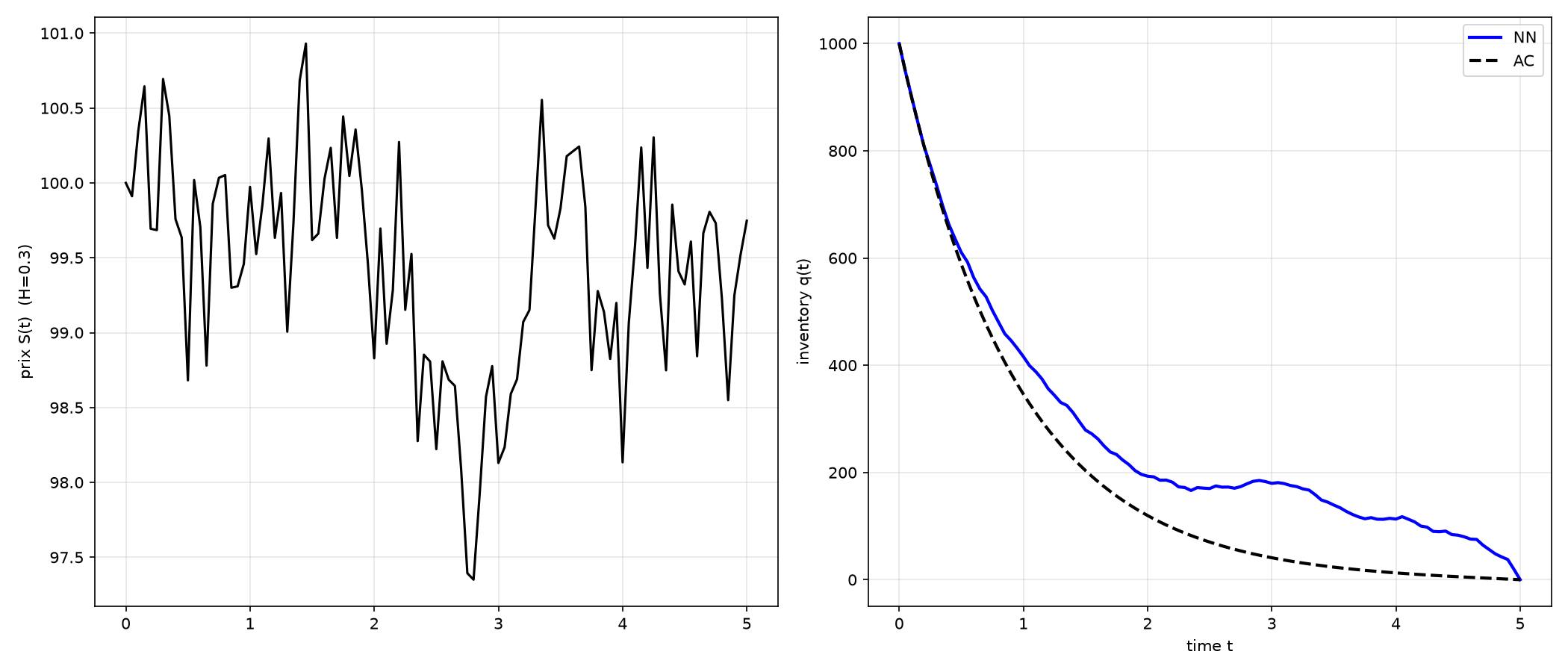}
    \captionsetup{width=0.8\textwidth}
    \caption{Median trajectory at $H=0.3$ ($\Pi_T^\mathrm{nn} - \Pi_T^\mathrm{ac} = +215$).}
    \label{fig:med03}
\end{figure}

\begin{figure}[htbp]
    \centering
    \includegraphics[width=0.85\linewidth]{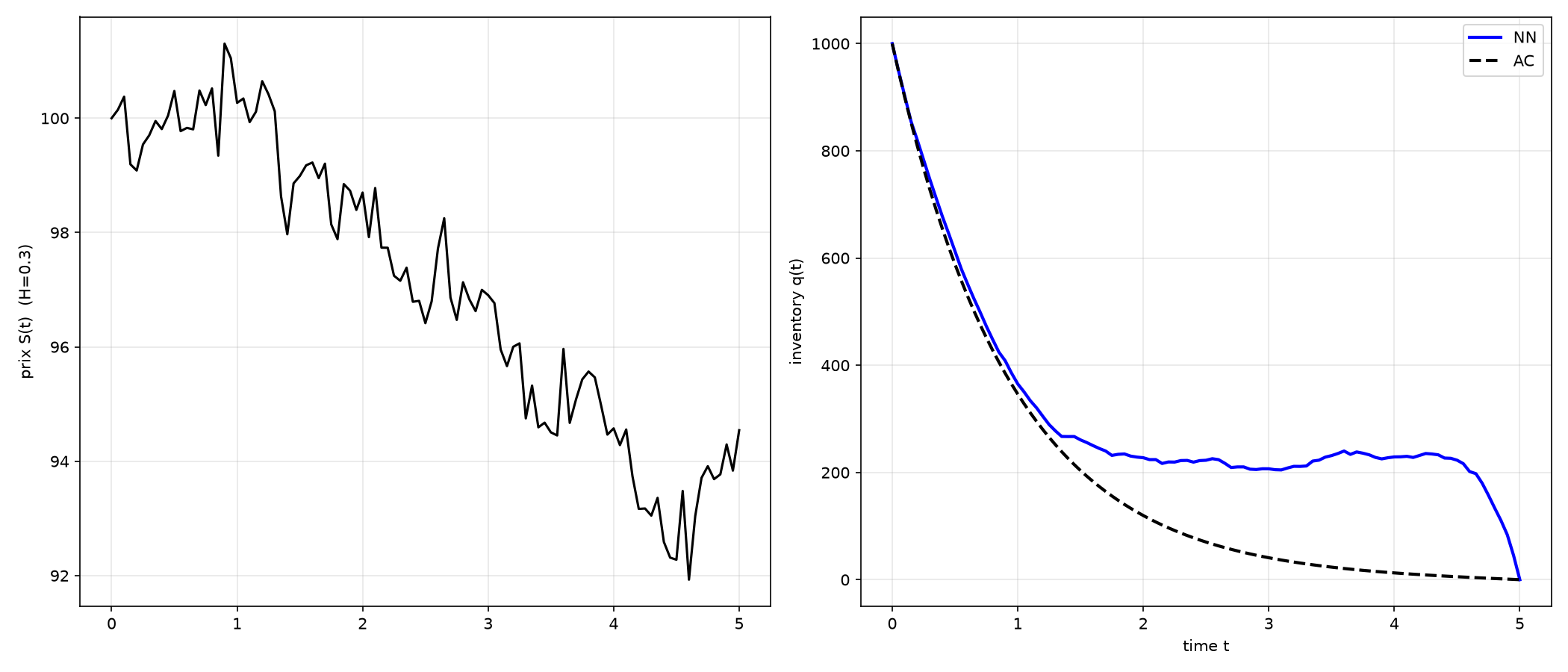}
    \captionsetup{width=0.8\textwidth}
    \caption{$1$st percentile trajectory at $H=0.3$ ($\Pi_T^\mathrm{nn} - \Pi_T^\mathrm{ac} = -863$).}
    \label{fig:perc1_03}
\end{figure}

The qualitative behavior is entirely different. On all three trajectories the network holds an inventory above the AC schedule, sometimes keeping a plateau of roughly $200$ shares for most of the horizon before liquidating in the final steps (Figures~\ref{fig:med03} \& \ref{fig:perc1_03}). Short positions rarely appear. This makes sense: with $H < 1/2$ the increments are negatively correlated, so a price that has just fallen is likely to bounce back, and holding inventory is a way of waiting for that reversal. Furthermore, since $\operatorname{Var}\{S_t - S_0\} = \sigma^2 t^{2H}$ is smaller for $H<1/2$ when $t \ge 1$ day, the price is less likely to explore extreme values, which makes delaying execution less risky than it would be at $H > 1/2$.

\begin{figure}[htbp]
    \centering
    \includegraphics[width=0.85\linewidth]{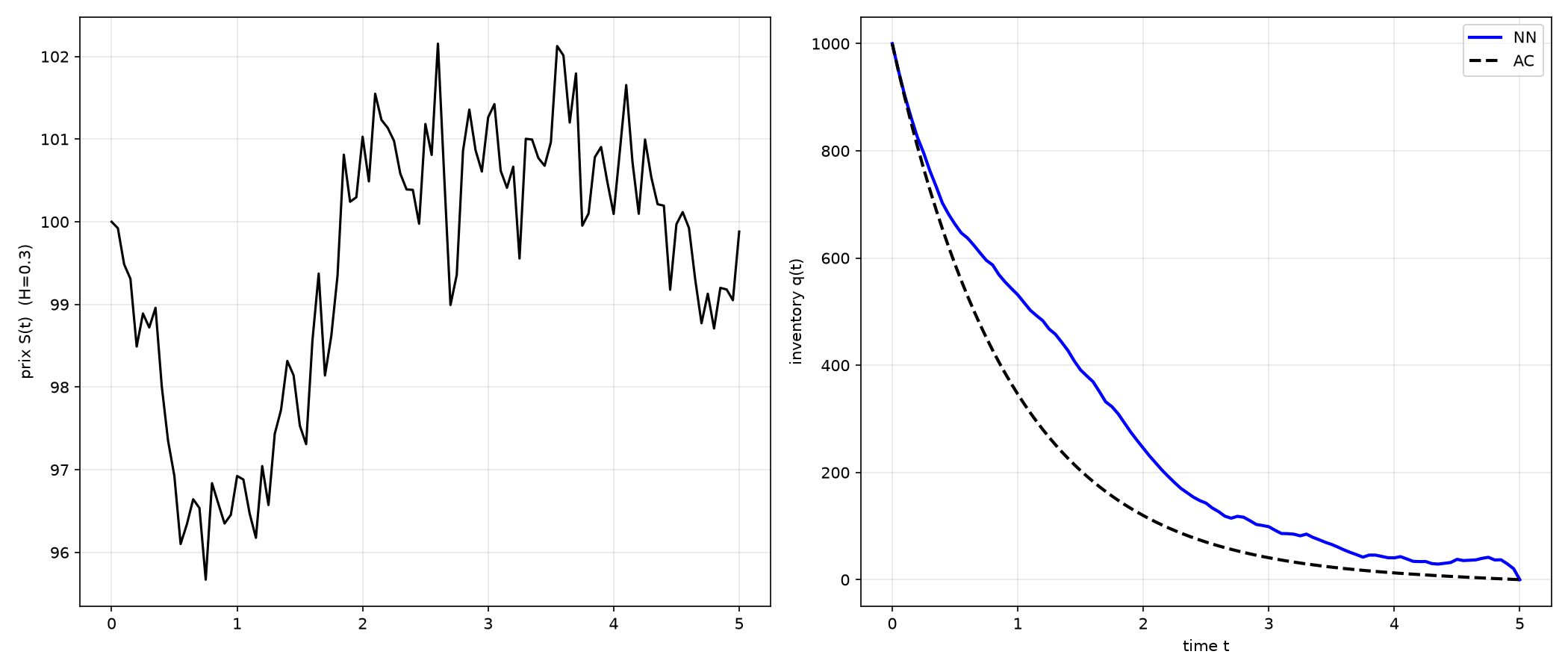}
    \captionsetup{width=0.8\textwidth}
    \caption{$99$th percentile trajectory at $H=0.3$ ($\Pi_T^\mathrm{nn} - \Pi_T^\mathrm{ac} = +766$).}
    \label{fig:perc99_03}
\end{figure}

Unlike the persistent case, in the median trajectory, the NN strategy already outperforms AC, as one can see in Figure \ref{fig:med03}. This is consistent with the larger aggregate gains reported for $H < 1/2$ in Section \ref{sub:perfToH}. Figure~\ref{fig:perc99_03} shows the favorable case: the price dips to $95.6$ early on, the network keeps a large inventory, and sells into the recovery above $100$. Figure~\ref{fig:perc1_03} shows the failure mode: the price declines throughout without ever reverting, and the inventory kept in reserve is liquidated at successively worse prices.

For {anti-persistent regimes} ($H < 1/2$), the out-performance is {obtained on most paths}.

The policy learns a genuinely different behavior on each side of $H = 1/2$: trend-following with shorting when increments are persistent, inventory-holding and mean-reversion capture when they are anti-persistent.

\subsection{Performances with regard to H}
\label{sub:perfToH}
\begin{figure}[htbp]
    \centering
    \includegraphics[width=0.85\textwidth]{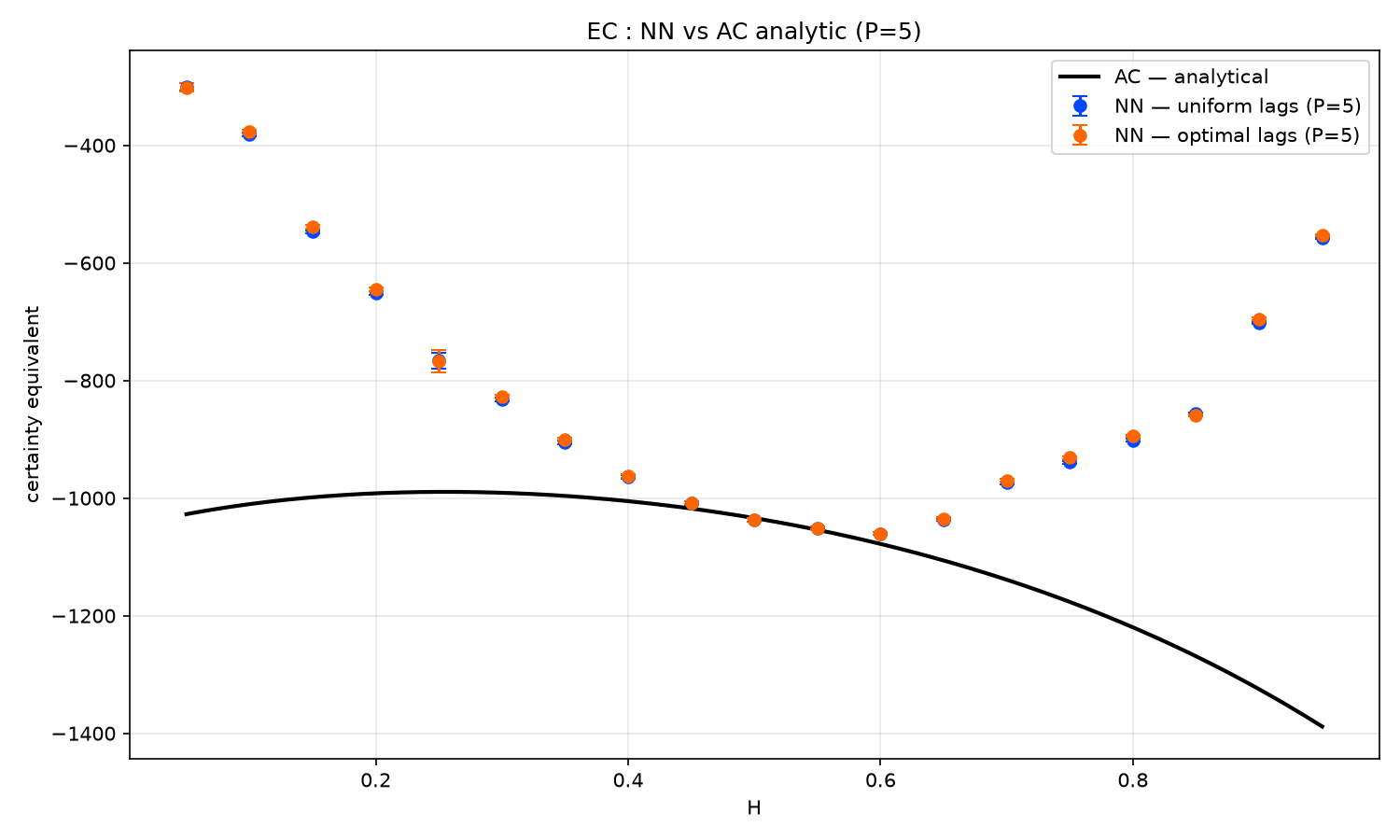}
    \captionsetup{width=0.7\textwidth}
    \caption{Certainty equivalent as a function of the Hurst exponent $H$: exact analytical value for the AC strategy versus the NN policy with $p=5$ (optimal or uniform) lagged returns.}
    \label{fig:ec_nn_vs_ac}
\end{figure}

Figure~\ref{fig:ec_nn_vs_ac} compares the certainty equivalent achieved by the learned dynamic policy for $p=5$ against the exact analytical certainty equivalent of the AC strategy, as a function of the Hurst exponent. First, at $H=1/2$ the network point sits on the AC curve, where it should be, because as mentioned before, in the martingale case, the optimal dynamic solution is in fact the deterministic one. Second, away from $H=1/2$ the network improves on AC, and the improvement grows as $H$ moves toward the extremes. Indeed, the more the increments are correlated, the more the conditional forecast $\hat{R}_{t,t+\tau}$ carries exploitable directional information \cite{BROUTY2024114543, garcinforecast}.

As in Section \ref{sub:perfACToH}, we observe an asymmetry for the NN and the AC performance, in favor of anti-persistent regimes, that we can again link to the variance evolution of $S_t-S_0$ with $H$. From Section \ref{sub:vizu}, we can also interpret this as a consequence of the trend-following behavior when $H>1/2$: the algorithm takes a bet that can be heavily punished by the exponential utility, whereas the holding strategy for $H<1/2$ protects us in most cases from the price trends.

We also compare both lag choices. Excluding the case $H = 1/2$, optimal lags yield a higher certainty equivalent in $15$ of the $18$ values of $H$. The difference $d = \operatorname{CE}_{\mathrm{opt}} - \operatorname{CE}_{\mathrm{unif}}$ is significant at the $5\%$ level in $10$ of them. The gain is modest compared to the improvement over AC, but consistent across both regimes, in line with the higher mean directional accuracy of optimal lags.

\subsection{Link to the mean-variance criterion}

We recall that via the cumulant generating function of $\Pi_T$, one founds that the CE equivalent is an approximation of the mean-variance criterion \eqref{meanvar} with $\widetilde\lambda = \lambda/2$. The learned policy can therefore improve on the AC baseline through two channels: increasing the expected payoff and reducing its variance.

\begin{figure}[htbp]
    \centering
    \includegraphics[width=0.85\textwidth]{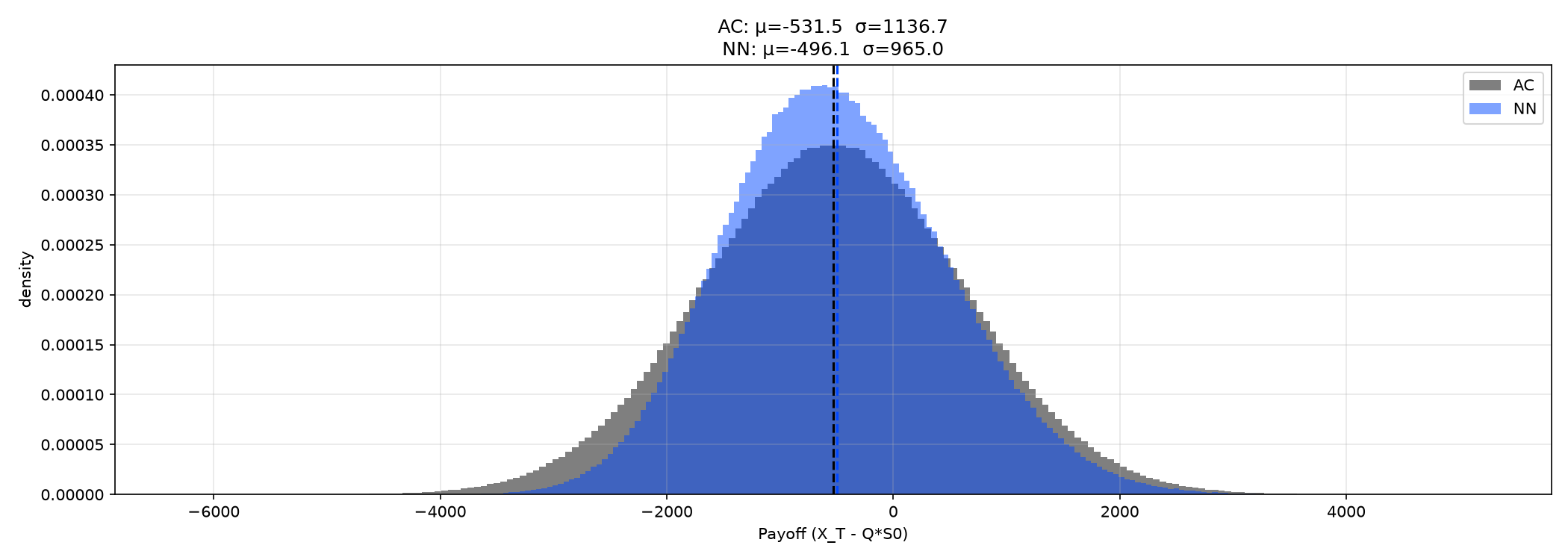}
    \captionsetup{width=0.65\textwidth}
    \caption{Distribution of the payoff $\Pi_T$
    for $H = 0.75$. The NN improves both the expected payoff and its
    variance. $\operatorname{MV}_\lambda^\mathrm{nn}=-1427.4$, $\operatorname{MV}_\lambda^\mathrm{ac}=-1823.5$ .}
    \label{fig:histo075}
\end{figure}

When $H$ is far from $1/2$, the forecast carries enough directional information for the policy to improve on both fronts. Figure~\ref{fig:histo075} illustrates this at $H = 0.75$.

Closer to $H = 1/2$, the forecast is weaker, and the trade-off becomes visible: the policy may sacrifice expected payoff in order to compress the tails of the distribution.
\begin{figure}[htbp]
    \centering
    \includegraphics[width=0.85\textwidth]{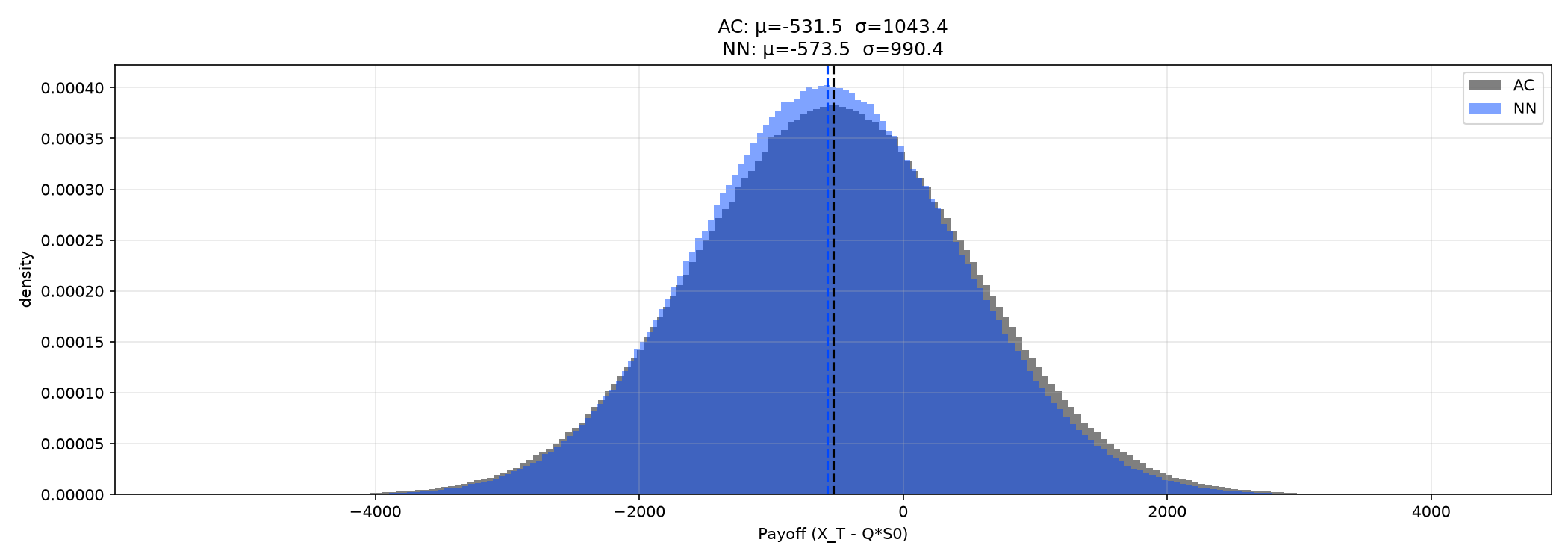}
    \captionsetup{width=0.65\textwidth}
    \caption{Distribution of the payoff $\Pi_T$
    for $H = 0.6$. The NN reduces variance at the cost of a lower mean
    payoff, and $\operatorname{MV}_\lambda^\mathrm{nn}=-1554.5$, $\operatorname{MV}_\lambda^\mathrm{ac}=-1620.3$ .}
    \label{fig:histo06}
\end{figure}
We notice that the mean-variance criterion improves on all values of $H \neq 1/2$.
\subsection{Robustness to parameter misspecification}

Our algorithm requires the Hurst exponent $H$ and the volatility $\sigma$ to simulate training trajectories and to compute the forecast $\hat{R}_{t,t+\tau}$. In practice, these parameters must be estimated from a finite sample of observed prices. A natural question is: how sensitive is the learned policy to estimation error in $(H, \sigma)$?

To answer this, we proceed as follows. We fix the true parameters at $H = 0.3$ and $\sigma = 1.5$ and we generate $M$ trajectories with these parameters. We estimate $(H,\sigma)$ using these $M$ trajectories, via the classical discrete variation estimator using the order-2 regression described in Section \ref{sub:estim}. We consider rather low values of $M$ in order to observe an estimation error of the parameters. 

We then train a full REINFORCE model using the estimates $(\widetilde{H}, \widetilde{\sigma})$ in place of the true values, both for the fBm simulation used in training and for the forecast computation. Finally, we evaluate each trained model on $n = 10^6$ test trajectories generated with the true parameters $(H, \sigma)$.

Table~\ref{tab:robustesse} reports the estimation results and the corresponding certainty equivalents. Even with a single trajectory ($M = 1$), where the estimation error is substantial ($\widetilde{H} = 0.361$ vs $H = 0.3$, $\widetilde{\sigma} = 1.78$ vs $\sigma = 1.5$), the learned policy still significantly outperforms the AC baseline. As $M$ increases and the estimates converge to the true values, the performance improves further, approaching the result obtained by the NN trained with the exact parameters.

\begin{table}[H]
    \centering
    \begin{tabular}{c|cc|cc}
        \toprule
        $M$ & $\widetilde{H}$ & $\widetilde{\sigma}$
            & $\mathrm{CE}^{\mathrm{nn}}$
            & $\mathrm{CE}^{\mathrm{ac}}$ \\
        \midrule
        $1$    & $0.361$ & $1.78$ & $-866.8 \pm 2.9$ & \multirow{5}{*}{$-990.9$} \\
        $10$   & $0.295$ & $1.52$ & $-846.8 \pm 2.9$ & \\
        $100$  & $0.294$ & $1.49$ & $-831.2 \pm 3.4$ & \\
        $1000$ & $0.297$ & $1.50$ & $-827.9 \pm 3.3$ & \\
        \text{True parameters}      & $0.300$   & $1.50$ & $-823.0 \pm 3.1$ & \\
        \bottomrule
    \end{tabular}
    \captionsetup{width=0.7\textwidth}
    \caption{Robustness to estimation error. Each row corresponds to a model trained with $(\widetilde{H}, \widetilde{\sigma})$ estimated from $M$ trajectories, and evaluates it on $n = 10^6$ test paths generated with the true parameters ($H = 0.3$, $\sigma = 1.5$). The AC baseline is independent of the estimation.}
    \label{tab:robustesse}
\end{table}

Two observations stand out. First, the policy is robust: even a $20\%$ relative error on $(H, \sigma)$ degrades the certainty equivalent by only $5\%$ compared to the model trained with the exact values, and all models outperform the AC baseline by a wide margin. Second, the performance improves monotonically with $M$, confirming that a moderate calibration sample is sufficient.

\section{Conclusion}

We developed a policy gradient algorithm that learns a liquidation strategy under fBm dynamics, conditioning its decisions on a single-scalar forecast of future returns built from $p$ past increments. The learned policy outperforms the AC baseline for every tested value of $H \neq 1/2$. It recovers it at $H = 1/2$, where the deterministic strategy is indeed optimal. The gain grows as $H$ moves away from $1/2$ and is markedly larger in the anti-persistent regime. A moderate number of lags suffices, most of the gain being already captured at $p = 5$, and the use of optimal lags \cite{garcinforecast} outperforms the use of regularly spaced ones in most of the tested cases.

The visualization of individual trajectories reveals that the network learns two qualitatively distinct behaviors: trend-following with short positions when increments are persistent, where the outperformance comes almost entirely from the tails of the payoff distribution, and inventory-holding to capture mean-reversion when increments are anti-persistent, where the gain is obtained on most paths. Although trained on the CARA criterion, the policy also improves the mean-variance criterion across all values of $H$, sometimes by diminishing the average payoff for reducing the variance. Finally, the method is robust to parameter misspecification: a $20\%$ error on $(H, \sigma)$ only costs about $5\%$ of the certainty equivalent.

However we still face some limitations. First, we need our price process to follow the fBm model. However, real market dynamics may be more complex, with potentially time-varying persistence, regime changes, and dependence structures that cannot be captured by a single, constant Hurst parameter~\cite{CCS,Garcin2017,Garcin2020}. Financial returns may exhibit as well pronounced time-varying volatility, including volatility clustering~\cite{Cont2001,Lux,ZZZ} and rough volatility patterns~\cite{GJR,GG,DF}. Second, despite our efforts, the learning procedure can still be chaotic when $H \to 0$ or $H \to 1$, since in these regimes the rewards fluctuate strongly.

\bibliographystyle{plain}
\bibliography{references}

@Article{almgren2000,
journal={Journal of risk},
author={Robert Almgren and Neil Chriss},
title={Optimal execution of portfolio transactions},
year={2000},
pages={5--39},
volume={3},
number={2}
}

@article{willingeretal,
  title={Stock market prices and long-range dependence},
  author={Willinger, Walter and Taqqu, Murad S and Teverovsky, Vadim},
  journal={Finance and stochastics},
  volume={3},
  number={1},
  pages={1--13},
  year={1999}
}

@Article{Guasoni2006,
journal={Mathematical finance},
author={Paolo Guasoni},
title={No Arbitrage Under Transaction Costs, With Fractional {B}rownian Motion And Beyond},
year={2006},
pages={569-582},
volume={16},
number={3}
}

@article{cheridito2003arbitrage,
  title={Arbitrage in fractional {B}rownian motion models},
  author={Cheridito, Patrick},
  journal={Finance and stochastics},
  volume={7},
  number={4},
  pages={533--553},
  year={2003}
}

@article{rogers,
author = {Rogers, Leonard Christopher Gordon},
title = {Arbitrage with Fractional {B}rownian Motion},
journal = {Mathematical finance},
volume = {7},
number = {1},
pages = {95--105},
year = {1997}
}

@article{famafrench,
 author = {Eugene F. Fama and Kenneth R. French},
 journal = {Journal of political economy},
 number = {2},
 pages = {246--273},
 title = {Permanent and Temporary Components of Stock Prices},
 volume = {96},
 year = {1988}
}

@article{lo1991,
 author = {Andrew W. Lo},
 journal = {Econometrica},
 number = {5},
 pages = {1279--1313},
 title = {Long-Term Memory in Stock Market Prices},
 volume = {59},
 year = {1991}
}

@article{guasoni2021high,
  title={High-frequency trading with fractional {B}rownian motion},
  author={Guasoni, Paolo and Mishura, Yuliya and R{\'a}sonyi, Mikl{\'o}s},
  journal={Finance and stochastics},
  volume={25},
  number={2},
  pages={277--310},
  year={2021}
}

@article{dolinsky2024,
  author  = {Yan Dolinsky and David Greenstein},
  title   = {A note on optimal liquidation with linear price impact},
  journal = {Modern stochastics: theory and applications},
  volume  = {12},
  number  = {2},
  pages   = {123--134},
  year    = {2025}
}

@article{mandelvanness,
author = {Mandelbrot, Benoit B. and Van Ness, John W.},
title = {Fractional {B}rownian Motions, Fractional Noises and Applications},
journal = {SIAM review},
volume = {10},
number = {4},
pages = {422-437},
year = {1968}
}

@article{bank2017hedging,
  title={Hedging with temporary price impact},
  author={Bank, Peter and Soner, H Mete and Vo{\ss}, Moritz},
  journal={Mathematics and financial economics},
  volume={11},
  number={2},
  pages={215--239},
  year={2017}
}

@article{timeconsistency,
author = {Vigna, Elena},
title = {ON TIME CONSISTENCY FOR MEAN-VARIANCE PORTFOLIO SELECTION},
journal = {International journal of theoretical and applied finance},
volume = {23},
number = {6},
pages = {1--22},
year = {2020}
}

@article{pedersen2017optimal,
  title={Optimal mean-variance portfolio selection},
  author={Pedersen, Jesper Lund and Peskir, Goran},
  journal={Mathematics and financial economics},
  volume={11},
  number={2},
  pages={137--160},
  year={2017}
}

@book{suttonbarto,
  author    = {Richard S. Sutton and Andrew G. Barto},
  title     = {Reinforcement learning: An introduction},
  edition   = {2nd},
  publisher = {MIT press},
  year      = {2018}
}

@article{garcinforecast,
  title={Forecasting with fractional {B}rownian motion: a financial perspective},
  author={Garcin, Matthieu},
  journal={Quantitative finance},
  volume={22},
  number={8},
  pages={1495--1512},
  year={2022}
}

@article{BROUTY2024114543,
title = {Fractal properties, information theory, and market efficiency},
journal = {Chaos, solitons \& fractals},
volume = {180},
pages = {114543},
year = {2024},
author = {Xavier Brouty and Matthieu Garcin}
}

@article{chanwood,
 author = {Andrew T. A. Wood and Grace Chan},
 journal = {Journal of computational and graphical statistics},
 number = {4},
 pages = {409--432},
 title = {Simulation of Stationary {G}aussian Processes in $[0,1]^d$},
 urldate = {2026-07-31},
 volume = {3},
 year = {1994}
}

@book{pham:hal-00401892,
  TITLE = {{Continuous-time stochastic control and optimization with financial applications}},
  AUTHOR = {Pham, Huy{\^e}n},
  PUBLISHER = {{Springer}},
  SERIES = {Stochastic modelling and applied probability n{\textdegree}61},
  PAGES = {xvii-232},
  YEAR = {2009}
}

@Article{coeurjollyestim,
journal={Statistical inference for stochastic processes},
author={Jean-François Coeurjolly},
title={Estimating the Parameters of a Fractional {B}rownian Motion by Discrete Variations of its Sample Paths},
year={2001},
pages={199-227},
volume={4},
number={2}
}

@article{gueant2015convex,
  title={A convex duality method for optimal liquidation with participation constraints},
  author={Gu{\'e}ant, Olivier and Lasry, Jean-Michel and Pu, Jiang},
  journal={Market microstructure and liquidity},
  volume={1},
  number={01},
  pages={1550002},
  year={2015}
}

@article{istas1997quadratic,
title = {Quadratic variations and estimation of the local {H}ölder index of a {G}aussian process},
journal = {Annales de l'institut Henri Poincare (B) probability and statistics},
volume = {33},
number = {4},
pages = {407-436},
year = {1997},
author = {Jacques Istas and Gabriel Lang}
}

@article{coeurjollysimul,
 title={Simulation and identification of the fractional {B}rownian motion: a bibliographical and comparative study},
 volume={5},
 number={7},
 journal={Journal of statistical software},
 author={Coeurjolly, Jean-Francois},
 year={2000},
 pages={1–53}
}

@article{Forde04032022,
author = {Martin Forde and Leandro Sánchez-Betancourt and Benjamin Smith},
title = {Optimal trade execution for {G}aussian signals with power-law resilience},
journal = {Quantitative finance},
volume = {22},
number = {3},
pages = {585--596},
year = {2022}
}

@article{han2013maximum,
  title={Maximum principle for general controlled systems driven by fractional {B}rownian motions},
  author={Han, Yuecai and Hu, Yaozhong and Song, Jian},
  journal={Applied mathematics \& optimization},
  volume={67},
  number={2},
  pages={279--322},
  year={2013}
}

@InProceedings{cont,
author="Cont, Rama",
editor="L{\'e}vy-V{\'e}hel, Jacques
and Lutton, Evelyne",
title="Long range dependence in financial markets",
booktitle="Fractals in engineering",
year="2005",
publisher="Springer London",
address="London",
pages="159--179",
}

@article{mandelbrot,
 author = {Benoit B. Mandelbrot},
 journal = {The review of economics and statistics},
 number = {3},
 pages = {225--236},
 title = {When Can Price be Arbitraged Efficiently? A Limit to the Validity of the Random Walk and Martingale Models},
 volume = {53},
 year = {1971}
}

@article{CCS,
  title={Time-dependent {H}urst exponent in financial time series},
  author={Carbone, Anna and Castelli, Giuliano and Stanley, H Eugene},
  journal={Physica A: statistical mechanics and its applications},
  volume={344},
  number={1-2},
  pages={267--271},
  year={2004}
}

@article{Cont2001,
  title={Empirical properties of asset returns: stylized facts and statistical issues},
  author={Cont, Rama},
  journal={Quantitative finance},
  volume={1},
  number={2},
  pages={223--236},
  year={2001}
}

@article{DF,
  title={Detecting rough volatility: a filtering approach},
  author={Damian, Camilla and Frey, R{\"u}diger},
  journal={Quantitative finance},
  volume={24},
  number={10},
  pages={1493--1508},
  year={2024}
}

@article{Garcin2017,
  title={Estimation of time-dependent {H}urst exponents with variational smoothing and application to forecasting foreign exchange rates},
  author={Garcin, Matthieu},
  journal={Physica A: statistical mechanics and its applications},
  volume={483},
  pages={462--479},
  year={2017}
}

@article{Garcin2020,
        author = {Garcin, Matthieu},
        year = {2020},
        title = {Fractal analysis of the multifractality of foreign exchange rates},
        journal = {Mathematical methods in economics and finance},
        volume = {13-14},
	number = {1},
        pages = {49--73}
}

@article{GG,
	  title={Long versus short time scales: the rough dilemma and beyond},
	  author={Garcin, Matthieu and Grasselli, Martino},
	  journal={Decisions in economics and finance},
	  volume={45},
	  number={1},
	  pages={257--278},
	  year={2022}
}

@article{GJR,
        author = {Gatheral, Jim and Jaisson, Thibault and Rosenbaum, Mathieu},
        year = {2018},
        title = {Volatility is rough},
        journal = {Quantitative finance},
        volume = {18},
	number = {6},
        pages = {933--949}
}

@article{Lux,
  title={The {M}arkov-switching multifractal model of asset returns: {GMM} estimation and linear forecasting of volatility},
  author={Lux, Thomas},
  journal={Journal of business \& economic statistics},
  volume={26},
  number={2},
  pages={194--210},
  year={2008}
}

@article{ZZZ,
  title={Equity markets volatility clustering: A multiscale analysis of intraday and overnight returns},
  author={Zhao, Xiaojun and Zhang, Na and Zhang, Yali and Xu, Chao and Shang, Pengjian},
  journal={Journal of empirical finance},
  volume={77},
  pages={101487},
  year={2024}
}
\end{document}